\documentclass[journal]{IEEEtran}

\usepackage{cite}
\ifCLASSINFOpdf
  \usepackage[pdftex]{graphicx}
  \graphicspath{{./fig/}}
  \DeclareGraphicsExtensions{.pdf,.jpeg,.png}
\else
  \usepackage[dvips]{graphicx}
  \graphicspath{{./fig/}}
  \DeclareGraphicsExtensions{.eps}
\fi
\usepackage{tikzexternal}
\tikzsetexternalprefix{tikz/}                   
\usepackage[cmex10]{amsmath}
\usepackage{amssymb}
\usepackage{amsthm}
\usepackage{mathtools}
\usepackage{stmaryrd}
\usepackage{mathrsfs}
\DeclareMathAlphabet{\mathscrbf}{OMS}{mdugm}{b}{n}
\usepackage{nicefrac}
\usepackage{enumerate}
\usepackage[algosection,ruled,vlined,titlenumbered,linesnumbered]{algorithm2e}
\Setnlskip{1ex}
\SetInd{.75ex}{.75ex}
\Setnlsty{tiny}{}{}
\SetKwComment{tgs}{\scriptsize/$\ast$\ }{\tiny\ $\ast$/}
\newcommand{\algcomment}[1]{\textsf{\tiny #1}}
\usepackage{array}
\ifCLASSOPTIONcompsoc
  \usepackage[caption=false,font=normalsize,labelfont=sf,textfont=sf]{subfig}
\else
  \usepackage[caption=false,font=footnotesize]{subfig}
\fi
\usepackage{dblfloatfix}
\usepackage{url}
\usepackage{hyperref}

\pdfpageattr{/Group <</S /Transparency /I true /CS /DeviceRGB>>} 

\newcommand{\ie}{i.e.,\ }
\newcommand{\eg}{e.g.,\ }
\newcommand{\defeq}{\ensuremath{\triangleq}}
\renewcommand{\vec}[1]{\ensuremath{\boldsymbol{#1}}}
\newcommand{\mat}[1]{\ensuremath{\boldsymbol{#1}}}
\newcommand{\q}{\ensuremath{q}}
\newcommand{\F}{\ensuremath{\mathbb{F}}}
\newcommand{\Fq}{\ensuremath{\F_{\q}}}
\newcommand{\N}{\ensuremath{\mathbb{N}}}
\newcommand{\REENCf}[1]{\ensuremath{\mathscr{R}_{#1}}}
\newcommand{\REENC}[2]{\ensuremath{\REENCf{#1}\left[#2\right]}}
\newcommand{\GRS}{\ensuremath{\mathcal{GRS}}}

\newcommand{\GRSparloc}{\ensuremath{\GRS_{\mathcal{A}, \mathcal{B}}\left(\Fq; n, k\right)}}
\newcommand{\Hw}[1]{\ensuremath{\mathrm{wt_\mathrm{H}}\left[#1\right]}}
\newcommand{\erasure}{\ensuremath{\vartimes}}
\renewcommand{\char}{\ensuremath{\mathop{\mathrm{char}}}}
\newcommand{\maxdeg}[2]{\ensuremath{d_{#1_{#2}}}}

{\theoremstyle{plain}
\newtheorem{theorem}{Theorem}
\newtheorem{lemma}{Lemma}

}

{\theoremstyle{definition}
\newtheorem{definition}{Definition}
\newtheorem{problem}{Problem}
}

{\theoremstyle{remark}
\newtheorem{example}{Example}
}

\begin{document}
\title{Prefactor Reduction of the Guruswami--Sudan Interpolation Step}
\author{Christian~Senger,~\IEEEmembership{Member,~IEEE}
\thanks{Christian Senger is with The Edward S.\ Rogers Sr.\ Department of Electrical and Computer Engineering, University of Toronto, 10 King's College Road (SFB540), Toronto, Ontario M5S 3G4, Canada. eMail: csenger@comm.utoronto.ca.}
}

\maketitle

\begin{abstract}
The concept of prefactors is considered in order to decrease the complexity of the Guruswami--Sudan interpolation step for generalized Reed--Solomon codes. It is shown that the well-known re-encoding projection due to K\"otter et al. \cite{koetter_ma_vardy_ahmed:2003} leads to one type of such prefactors. The new type of Sierpinski prefactors is introduced. The latter are based on the fact that many binomial coefficients in the Hasse derivative associated with the Guruswami--Sudan interpolation step are zero modulo the base field characteristic. It is shown that both types of prefactors can be combined and how arbitrary prefactors can be used to derive a \emph{reduced} Guruswami--Sudan interpolation step.
\end{abstract}


\IEEEpeerreviewmaketitle

\section{Introduction}

\IEEEPARstart{D}{ecoding} of generalized Reed--Solomon codes comes in multiple flavors. Among the most widely-deployed techniques for decoding up to $\left\lfloor\nicefrac{(n-k)}{2}\right\rfloor$ errors, where $n$ is the length and $k$ is the dimension of the code, is decoding based on linear shift register synthesis with the Berlekamp--Massey algorithm \cite{massey:1969}. Other techniques utilize the extended Euclidean algorithm \cite{sugiyama_kasahara_hirasawa_namekawa:1975, gao:2002}, Newton interpolation \cite{sorger:1993}, and bivariate polynomial interpolation \cite{welch_berlekamp:1986, gemmell_sudan:1992}. Apart from implementational speedups, \eg divide \& conquer techniques that allow to implement the Berlekamp--Massey algorithm in $\mathcal{O}[n\log[n]]$ as proposed in \cite{blahut:1983}, all these algorithms are in $\mathcal{O}[n^2]$.

Decoding beyond $\left\lfloor\nicefrac{(n-k)}{2}\right\rfloor$ is much more complicated and efficient (polynomial-time) decoding algorithms for this case have been known only after the invention of the Sudan algorithm in 1997 \cite{sudan:1997} and the Guruswami--Sudan algorithm in 1999 \cite{guruswami_sudan:1999}. The latter can correct up to $n-\sqrt{nk}$ errors.

As we will see shortly, the Guruswami--Sudan algorithm consists of two steps where one of them --- an interpolation problem --- is computationally more involved than the other. Extensive efforts to speed up the interpolation problem include exploiting structured matrices \cite{olshevsky_shokrollahi:2003, zeh_gentner_augot:2011} or the underlying algebraic structure \cite{koetter:1996, alekhnovich:2002, trifonov:2010}. The technique of re-encoding was introduced in \cite{gross_kschischang_koetter_gulak:2005, koetter_ma_vardy:2011}; it allows one to make predictions about the structure of the solutions of the interpolation problem and then to exploit this knowledge in order to reduce the problem. Our contribution in this paper is to show how the size of the interpolation problem can in many cases be reduced beyond re-encoding, utilizing a simple property of the generalized Reed--Solomon code's base field. In order to provide a framework that unites re-encoding, our proposal, and the combination of both, we introduce the general notion of a-priori known \emph{prefactors} and show how such factors can lead to a reduction of the interpolation problem.

The rest of this paper is organized as follows. Section~\ref{sec:prelim} covers generalized Reed--Solomon codes as well as the Guruswami--Sudan algorithm with its interpolation and factorization steps. The concept of re-encoding is explained in Section~\ref{sec:reenc} and a theorem delivers the associated \emph{re-encoding} prefactors. Section~\ref{sec:sierpinski} represents the main part of the paper. It introduces the new type of \emph{Sierpinski} prefactors and gives results about their existence and properties. Section~\ref{sec:factors} shows how prefactors of both types can be combined and how they can be used to diminish the size of the interpolation problem. One of our main results is given in this section, \ie the reduced interpolation step given by Problem~\ref{prob:interpolation_reduced}. The paper is wrapped up in Section~\ref{sec:conclusion} and two useful algorithms are provided in an appendix.

\section{Preliminaries}\label{sec:prelim}

\begin{definition}\label{def:grs}
For a prime power $q$ and $n,k\in\N\setminus\{0\}$ with $k\leq n\leq q-1$ let $\mathcal{A}=\{\alpha_0, \ldots, \alpha_{n-1}\}$ be an ordered set of distinct elements (\emph{code locators}) from the finite field $\Fq$ and let $\mathcal{B}=\{\beta_0, \ldots, \beta_{n-1}\}$ be an ordered set of nonzero (not necessarily distinct) elements (\emph{column multipliers}) from $\Fq$. Then the set of vectors
\begin{multline*}
 \GRSparloc\defeq\left\{\left(\beta_0 u(\alpha_0), \ldots, \beta_{n-1}u(\alpha_{n-1})\right)\right.:\\%
 \left.\vphantom{\left(\beta_0 u(\alpha_0), \ldots, \beta_{n-1}u(\alpha_{n-1})\right)}u(x)\in\Fq[x],\deg[u(x)]<k\right\}
\end{multline*}
is a \emph{generalized Reed--Solomon (GRS) code} \cite{reed_solomon:1960,delsarte:1975}.
\end{definition}

Since we frequently assume that \emph{base field} $\Fq$, \emph{code length} $n$, and \emph{code dimension} $k$ as well as code locators and column multipliers are fixed, we write $\GRS$ for $\GRSparloc$ whenever it is convenient. GRS codes fulfill the Singleton bound with equality, \ie their \emph{minimum distance} is $d=n-k+1$. $\GRS$ is a linear subspace of $\Fq^n$, and thus it is a linear code. Note that \emph{conventional Reed--Solomon (RS)} codes are special cases of GRS codes with $\mathcal{A}=\{1, \alpha, \ldots, \alpha^{n-1}\}$ and $\mathcal{B}=\{1, \alpha^{b}, \ldots, \alpha^{b(n-1)}\}$, where $\alpha\in\Fq$ has order $n$ and $b\in\N$ \cite{roth:2006}.

The state-of-the-art of decoding GRS codes is the \emph{Guruswami--Sudan algorithm (GSA)} \cite{guruswami_sudan:1999}.\footnote{This statement neglects the Wu algorithm \cite{wu:2008}, which achieves the same error-correcting capabilities as the GSA and is based on rational interpolation.} It can be divided into two steps: the interpolation step (Problem~\ref{prob:interpolation}) and the factorization step (Problem~\ref{prob:factorization}). Our focus here is on the interpolation step, which is computationally more involved.

Let $\vec{c}\in\GRS$ be a codeword, $\vec{e}\in\Fq^n$ be an error vector of Hamming weight $\Hw{\vec{e}}=\varepsilon$, and $\vec{y}=\vec{c}+\vec{e}$ be the corresponding received vector obtained from the transmission channel. Furthermore, let $\mathcal{I}\defeq\{0, \ldots, n-1\}$ and let $r,\ell\in\N\setminus\{0\}$ be two parameters of the GSA with $r\leq \ell$. With any set $\mathcal{S}\subseteq\mathcal{I}$ we associate the polynomial
\begin{equation}\label{eqn:Px}
  P_\mathcal{S}(x)\defeq\prod_{i\in\mathcal{S}}(x-\alpha_i)
\end{equation}
with $P_\mathcal{S}=1$ if $\mathcal{S}=\emptyset$. Here, $\alpha\in\Fq$ is a primitive element. 

\begin{problem}[GSA Interpolation Step]\label{prob:interpolation}
  Given a received vector $\vec{y}$ and $\varepsilon_0\in\N$, find a nonzero bivariate polynomial $Q(x, z)=Q_0(x)+Q_1(x)z+\cdots +Q_\ell(x)z^\ell\in\Fq[x, z]$ such that 
  \begin{equation*}
    \deg\left[Q_\nu(x)\right]\leq r(n-\varepsilon_0)-\nu(k-1)-1\defeq \maxdeg{Q}{\nu}
  \end{equation*}
  for $\nu=0, \ldots, \ell$ and\vspace{-0.3cm}
  \begin{multline}\label{eqn:hasse}
    \forall i\in\mathcal{I}\;\forall s,t\in\N:s+t<r\;\text{and}\\
    \sum_{\nu=t}^{\ell}\binom{\nu}{t}z^{\nu-t}\sum_{\mu=s}^{\maxdeg{Q}{\nu}}
    \binom{\mu}{s}x^{\mu-s}Q_{\nu, \mu}%
    \bigg|_{(x, z)=\left(\alpha^{-i}, y_i\right)}=0,
  \end{multline}
  where $Q_\nu(x)=\sum_{\mu=0}^{\maxdeg{Q}{\nu}} Q_{\nu, \mu}x^\mu$.
\end{problem}

The nested sum in \eqref{eqn:hasse} is called the $(s, t)$th \emph{mixed partial Hasse derivative} \cite{hasse:1936} of $Q(x, z)$. The condition that all $(s, t)$th Hasse derivatives with $s+t<r$ evaluate to zero for all tuples $(\alpha^{-i}, y_i)$, $i\in\mathcal{I}$, means that these tuples are zeros of multiplicity $r$ of $Q(x, z)$. For that reason, we refer to the parameter $r$ as the \emph{multiplicity} of the GSA.

A straightforward analysis shows that the homogeneous linear system of equations associated with Problem~\ref{prob:interpolation} has $n\binom{r+1}{2}$ equations and $\sum_{\nu=0}^{\ell}\left(\maxdeg{Q}{\nu}+1\right)$ unknowns. Both numbers are exceedingly large even for short GRS codes and intermediate parameters $r$ and $\ell$. It can be shown that the linear system has a nonzero solution $Q(x, z)$ (\ie it has more equations than unknowns) as long as
\begin{equation*}
  \varepsilon<\frac{n(2\ell-r+1)}{2(\ell+1)}-\frac{\ell(k-1)}{2r}\defeq\varepsilon_0.
\end{equation*}

Naively solving the linear system with Gaussian elimination in order to obtain a solution is in $\mathcal{O}\left[\ell^6 n^3\right]$ since $\binom{r+1}{2}=\nicefrac{(r+1)!}{2(r-1)!}=\nicefrac{r(r+1)}{2}$ and $r\leq\ell$. One of the fastest methods to solve the interpolation step is K\"otter interpolation \cite{koetter:1996}, which is in $\mathcal{O}\left[\ell^4 n^2\right]$. A fast algorithm that actually solves the homogeneous linear system  was provided in \cite{zeh_gentner_bossert:2009} and is based on results from \cite{roth_ruckenstein:2000}.

Without loss of generality we assume in the following that the columns of the coefficient matrix (from left to right) are associated with the unknown coefficients
\begin{equation*}
  Q_{0, 0}, \ldots, Q_{0, \maxdeg{Q}{0}}, Q_{1, 0}, \ldots, Q_{1, \maxdeg{Q}{1}}, \ldots, Q_{\ell, 0}, \ldots, Q_{\ell, \maxdeg{Q}{\ell}}.
\end{equation*}
Such a coefficient matrix can be set up using Algorithm~\ref{alg:GSA_matrix} in the appendix. The particular ordering of the coefficients will become important in Section~\ref{sec:factors} where we apply our results in order to reduce the coefficient matrix.

\begin{problem}[GSA Factorization Step]\label{prob:factorization}
  Given a solution $Q(x, z)$ of Problem~\ref{prob:interpolation}, find all factors of the form $z-w(x)$ with $w(x)\in\Fq[x]$, $\deg\left[w(x)\right]<k$.
\end{problem}

Let us collect all such $w(x)$ in a set $\mathcal{W}$. From $\mathcal{W}$ we can calculate the \emph{result list}
\begin{equation*}
 \mathcal{L}\defeq\left\{\left(\beta_0 w(\alpha_0), \ldots, \beta_{n-1}w(\alpha_{n-1})\right):w(x)\in\mathcal{W}\right\}.
\end{equation*}
Obviously, $\mathcal{L}\subseteq\GRS$ and $\vert\mathcal{L}\vert\leq\ell$. Due to the latter fact, we refer to the parameter $\ell$ as the \emph{list size}. Note that all codewords $\vec{c}'\in\mathcal{L}$ with $\Hw{\vec{y}-\vec{c}'}>\varepsilon_0$ can be discarded. It is proven in \cite{guruswami_sudan:1999} that $\vec{c}\in\mathcal{L}$ if $\Hw{\vec{y}-\vec{c}}=\Hw{\vec{e}}\leq\varepsilon_0$.

Problem~\ref{prob:factorization} can be solved with time complexity in $\mathcal{O}\left[\ell \log\log[\ell]n^2\right]$ using a technique from \cite{roth_ruckenstein:2000}, but this is not the focus of this paper.

It follows from  the exposition of the Welch--Berlekamp algorithm in \cite{gemmell_sudan:1992} and the interpretation of Justesen and H\o holdt in \cite[Sections~5.2 and 12.2]{justesen_hoholdt:2004} that the GSA simplifies to the Sudan algorithm \cite{sudan:1997} if we restrict the multiplicity to $r=1$ and that it further simplifies to the Welch--Berlekamp algorithm \cite{welch_berlekamp:1986} if we additionally restrict the list size to $\ell=1$.

\section{The Re-Encoding Projection}\label{sec:reenc}

The re-encoding projection in the context of the GSA was first introduced in \cite{koetter_ma_vardy_ahmed:2003} and later elaborated for application with interpolation algorithms in \cite{gross_kschischang_koetter_gulak:2005,ma:2007,koetter_ma_vardy:2011}. Its key idea is to project the received vector $\vec{y}\in\Fq^n$ onto a subspace of $\Fq^n$ with zero components at positions indexed by $\mathcal{J}\subseteq\mathcal{I}$, $\vert\mathcal{J}\vert=k$. Not surprisingly, it turns out that this allows to skip $k$ of the interpolation points $(\alpha^{-i}, y_i)$ in \eqref{eqn:hasse} and thus to reduce the size of the interpolation Problem~\ref{prob:interpolation} from $n$ to $n-k$ points.

\begin{definition}
The \emph{re-encoding projection} with respect to $\mathcal{J}\subseteq\mathcal{I}$, $\vert\mathcal{J}\vert=k$, for $\GRSparloc$ is defined as the linear map
\begin{equation*}
  \REENCf{\mathcal{J}} ~:~ \left\{\begin{array}{rcl} \Fq^n &\to &\Fq^n\\
  \vec{v} &\mapsto &\vec{v}+\widetilde{\vec{c}}\end{array}\right.,
\end{equation*}
where $\widetilde{\vec{c}}\in\GRS$, such that $\forall i\in\mathcal{J}:\widetilde{c}_i=-v_i$.  
\end{definition}

The re-encoding projection annihilates the components $v_i$, $i\in\mathcal{J}$, and sets them to zero. It is injective since GRS codes fulfill the Singleton bound with equality and thus any codeword $\widetilde{\vec{c}}\in\GRSparloc$ is uniquely determined by the $k$ components $\widetilde{c}_i$, $i\in\mathcal{J}$. The remaining $n-k$ components $\widetilde{c}_i$, $i\in\mathcal{I}\setminus\mathcal{J}$, can be efficiently found using an erasures-only decoder. $\REENCf{\mathcal{J}}$ is indeed a projection, \ie it is idempotent. This can be seen by the fact that all components of $\REENC{\mathcal{J}}{\vec{v}}=\vec{v}+\widetilde{\vec{c}}_1$ at positions $i\in\mathcal{J}$ are zero by definition. But then $\REENC{\mathcal{J}}{\REENC{\mathcal{J}}{\vec{v}}}=\vec{v}+\widetilde{\vec{c}}_1+\widetilde{\vec{c}}_2$, where $\widetilde{\vec{c}}_2\in\GRS$ is zero at the $k$ positions in $\mathcal{J}$, which is only possible if it is the all-zero codeword.

\begin{example}\label{ex:reenc1}
Consider the conventional RS code $\GRS_{\mathcal{A}, \mathcal{B}}(\F_{11}; 11, 5)$ with $\mathcal{A}=\{\alpha^i:i=0, \ldots, 10\}=\{1, 2, 4, 8, 5, 10, 9, 7, 3, 6\}$ (primitive element $\alpha=2$) and $\mathcal{B}=\{1, 1, 1, 1, 1, 1, 1, 1, 1, 1\}$. For this code, the GSA with multiplicity $r=2$ and list size $\ell=3$ can correct up to $\varepsilon_0=3$ errors. Assume the following composition of the received vector:
\begin{equation*}
  \setlength{\arraycolsep}{0.1em}
  \begin{array}{rrrrrrrrrrrl}
      & ( \phantom{1}5, &  3, &  8, & 10, &  7, &  8, &  4, &  5, &  6, &  4) & =\vec{c}\\
    + & ( \phantom{1}0, &  0, &  0, &  0, &  0, &  2, &  1, &  0, &  7, &  0) & =\vec{e}\\
    \hline
    = & ( \phantom{1}5, &  3, &  8, & 10, &  7, & 10, &  5, &  5, &  2, &  4) & = \vec{y}
  \end{array}
\end{equation*}
Assume further $\mathcal{J}=\{5, 6, 7, 8, 9\}$, \ie when we apply the re-encoding projection to $\vec{y}$ we obtain $(\erasure, \erasure, \erasure, \erasure, \erasure, 1, 6, 6, 9, 7)$ and need to correct the $\erasure$ using an erasures-only decoder. The (unique) result of this procedure is $\widetilde{\vec{c}}=(3, 3, 4, 10, 5, 1, 6, 6, 9, 7)$. With this, the re-encoding projection of $\vec{y}$ becomes:
\begin{equation*}
  \setlength{\arraycolsep}{0.1em}
  \begin{array}{rrrrrrrrrrrl}
      & ( \phantom{1}5, &  3, &  8, & 10, &  7, & 10, &  5, &  5, &  2, &  4) & = \vec{y}\\
    + & ( \phantom{1}3, &  3, &  4, & 10, &  5, &  1, &  6, &  6, &  9, &  7) & = \widetilde{\vec{c}}\\
    \hline
    = & ( \phantom{1}8, &  6, &  1, &  9, &  1, &  0, &  0, &  0, &  0, &  0) & = \REENCf{\mathcal{J}}[\vec{y}]
  \end{array}
\end{equation*}
But then $\REENCf{\mathcal{J}}[\vec{y}]=\vec{c}+\widetilde{\vec{c}}+\vec{e}$ with $\vec{c}+\widetilde{\vec{c}}\in\GRS$, hence we have a new received vector with (as we will see) favorable properties but still the original error vector:
\begin{equation*}
  \setlength{\arraycolsep}{0.1em}
  \begin{array}{rrrrrrrrrrrl}
      & ( \phantom{1}8, &  6, &  1, &  9, &  1, &  9, & 10, &  0, &  4, &  0) & =\vec{c}+\widetilde{\vec{c}}\\
    + & ( \phantom{1}0, &  0, &  0, &  0, &  0, &  2, &  1, &  0, &  7, &  0) & =\vec{e}\\
    \hline
    = & ( \phantom{1}8, &  6, &  1, &  9, &  1, &  0, &  0, &  0, &  0, &  0) & = \REENCf{\mathcal{J}}[\vec{y}]
  \end{array}
\end{equation*}
\end{example}

The example shows how the re-encoding projection can be used in order to simplify the decoding: First, the received vector $\vec{y}$ is projected, which yields a new received vector $\REENCf{\mathcal{J}}[\vec{y}]$ with (at least) $k$ zero components. Then, the GSA is applied to the new received vector in order to obtain the error vector $\vec{e}$. This can be done more efficiently than decoding of $\vec{y}$ due to the designed properties of $\REENCf{\mathcal{J}}[\vec{y}]$. Finally, the error vector together with the original received vector can be used to calculate the transmitted codeword $\vec{c}$.

The following theorem shows that decoding of $\REENCf{\mathcal{J}}[\vec{y}]$ with the GSA induces strong structural properties on the bivariate result polynomial $Q(x, z)$ of the interpolation step or, more precisely, its constituent univariate polynomials $Q_\nu(x)$, $\nu=0, \ldots, \ell$. It will become clear in Section~\ref{sec:factors} how this can be exploited in order to reduce the size of the interpolation Problem~\ref{prob:interpolation}.

\begin{theorem}\label{thm:reenc_factors}
Let $\GRSparloc$ be a GRS code, $\mathcal{J}\subseteq\mathcal{I}$ with $\vert\mathcal{J}\vert=k$, and $r,\ell$ such that the GSA can correct at most $\varepsilon_0$ errors. Let further $\vec{c}\in\GRS$, $\vec{e}\in\Fq^n$ with $\Hw{\vec{e}}\leq\varepsilon_0$ and $\vec{y}=\vec{c}+\vec{e}$. When the GSA is applied to $\REENCf{\mathcal{J}}[\vec{y}]$ it yields a bivariate result polynomial $Q(x, z)=Q_0(x)+Q_1(x)z+\cdots +Q_\ell(x)z^\ell\in\Fq[x, z]$ whose constituent univariate polynomials $Q_\nu(x)$, $\nu=0, \ldots, r-1$, can be factored as
\begin{equation}\label{eqn:reenc_factors}
  Q_\nu(x)=U_\nu(x)P_\mathcal{J}(x)^{r-\nu},
\end{equation}
where $\deg[U_\nu(x)]\leq \maxdeg{Q}{\nu}-k(r-\nu)\defeq\maxdeg{U}{\nu}$.
\end{theorem}

We emphasize that the \emph{re-encoding prefactors} $P_\mathcal{J}(x)^{r-\nu}$ in \eqref{eqn:reenc_factors} are fixed a-priori, \ie they do not depend on the received vector $\vec{y}$. This fact will be exploited in Section~\ref{sec:factors}, it basically allows to work with the quotient polynomials $U_\nu(x)$ instead of the full polynomials $Q_\nu(x)$. Note that we required $r\leq\ell$ in Section~\ref{sec:prelim}, hence there generally are polynomials $Q_\nu(x)$, $\nu=r, \ldots, \ell$ that cannot be factored.

\begin{IEEEproof}[Proof of Theorem~\ref{thm:reenc_factors}]
Assume that $Q(x, z)$ is a solution of Problem~\ref{prob:interpolation} for the projected received vector $\REENCf{\mathcal{J}}[\vec{y}]$. That is, the coefficients of $Q(x, z)$ fulfill \eqref{eqn:hasse}. If we restrict \eqref{eqn:hasse} to $i\in\mathcal{J}\subseteq\mathcal{I}$ we obtain
\begin{multline}\label{eqn:reenc_hasse1}
  \forall i\in\mathcal{J}\;\forall s,t\in\N:s+t<r\;\text{and}\\
  \sum_{\nu=t}^{\ell}\binom{\nu}{t}0^{\nu-t}\sum_{\mu=s}^{\maxdeg{Q}{\nu}}
  \binom{\mu}{s}x^{\mu-s}Q_{\nu, \mu}%
  \bigg|_{x=\alpha^{-i}}=0,
\end{multline}
since $\forall i\in\mathcal{J}:y_i=0$. The factor $0^{\nu-t}$ annihilates all summands of the outer sum except the one for $\nu=t$, which yields $0^{\nu-t}=0^0=1$. But then \eqref{eqn:reenc_hasse1} becomes
\begin{multline}\label{eqn:reenc_hasse2}
  \forall i\in\mathcal{J}\;\forall s,t\in\N:s<r-t\;\text{and}\\
  \sum_{\mu=s}^{\maxdeg{Q}{t}}
  \binom{\mu}{s}x^{\mu-s}Q_{t, \mu}%
  \bigg|_{x=\alpha^{-i}}=0,
\end{multline}
which means that the $0$th to $r-t-1$th Hasse derivatives of the $Q_t(x)$, $t=0, \ldots, r-1$, evaluate to zero at all $\alpha^{-i}$, $i\in\mathcal{J}$. But then the $\alpha^{-i}$, $i\in\mathcal{J}$, are roots of multiplicity $r-t$ of $Q_t(x)$, which (after replacing $t$ by $\nu$) proves our argument. The bound on the degrees of the $U_\nu(x)$ is easy to see by comparing the degrees of the involved polynomials.
\end{IEEEproof}

\begin{example}\label{ex:reenc2}
For the setting of Example~\ref{ex:reenc1}, the interpolation step of the GSA could result in the bivariate polynomial
\begin{equation*}
  Q(x, z)=Q_0(x)+Q_1(x)z+Q_2(x)z^2+Q_3(x)z^3
\end{equation*}
with univariate constituent polynomials
\begin{IEEEeqnarray*}{rcl}
  Q_0(x) & = & 6x^{13} + 2x^{12} + 6x^{11} + x^{10} + 2x^9 + 7x^8 + 2x^7\\
         &   & +\: 4x^6 + 2x^5 + 7x^4 + 3x^3 + 3x^2 + 8x + 5\\
  Q_1(x) & = & 7x^9 + x^8 + 9x^7 + 5x^4 + 6x^3 + 4x^2 + 1\\
  Q_2(x) & = & 2x^5 + 10x^4 + z^2 + 2x + 6\\
  Q_3(x) & = & 1.
\end{IEEEeqnarray*}
From $\mathcal{J}=\{5, 6, 7, 8, 9\}$ and \eqref{eqn:Px} we obtain
\begin{align*}
  P_\mathcal{J}(x) &=(x-\alpha^{-5})(x-\alpha^{-6})(x-\alpha^{-7})(x-\alpha^{-8})(x-\alpha^{-9})\\
       &=x^5 + 4x^4 + 8x^3 + 2x^2 + 9x + 1
\end{align*}
and it is easy to verify the factorizations
\begin{align*}
  Q_0(x) &= (\underbrace{6x^3 + 9x^2 + 6x + 5}_{=U_0(x)})P_\mathcal{J}(x)^2\\
  Q_1(x) &= (\underbrace{7x^4 + 6x^3 + 6x^2 + 2x + 1}_{=U_1(x)})P_\mathcal{J}(x)
\end{align*}
that are given by Theorem~\ref{thm:reenc_factors}.
\end{example}

We have seen in this section that the re-encoding projection can be used in order to obtain structured solutions, \ie bivariate polynomials $Q(x, z)$, of Problem~\ref{prob:interpolation}. The computational overhead of re-encoding is negligible, as it is confined to a single erasures-only decoding step for a vector with $n-k$ erased symbols.

\section{Sierpinski Prefactors}\label{sec:sierpinski}

In this section, we introduce a new technique that results in structured solutions of Problem~\ref{prob:interpolation}. In contrast to the re-encoding projection, this approach does \emph{not} require any additional computations, it simply exploits basic properties of the GRS code's base field $\Fq$. The main idea is to exploit the fact that many of the binomial coefficients in \eqref{eqn:hasse} are zero modulo the characteristic of $\Fq$.

\begin{figure*}[!t]
\centering
\subfloat[Sierpinski triangle $\mathfrak{S}_2$, $\binom{\nu}{t}\bmod 2$, $t_0=15$ is a zero column for $\ell\in\{16, \ldots, 30\}$ and $t_1=7$ is a zero column with resolvable spoiler $\binom{15}{7}$ for $\ell\in\{16, \ldots, 22\}$.]{
  \begin{tikzpicture}[scale=\pascalscale,every node/.style={transform shape}]    
  \end{tikzpicture}
\label{fig:sierpinski2}}
\hfil
\subfloat[Sierpinski triangle $\mathfrak{S}_3$, $\binom{\nu}{t}\bmod 3$, $t_0=8$ is a zero column for $\ell\in\{9, \ldots, 16\}$ and $t_1=7$ is a zero column with resolvable spoiler $\binom{8}{7}$ for $\ell\in\{9, \ldots, 15\}$.]{
  \begin{tikzpicture}[scale=\pascalscale,every node/.style={transform shape}]    
  \end{tikzpicture}
\label{fig:sierpinski3}}
\caption{Two instances of the first $31$ rows of the left-aligned Pascal triangle, where the binomial coefficients are calculated modulo $2$ and $3$, respectively, \ie the Sierpinski triangles $\mathfrak{S}_2$ and $\mathfrak{S}_3$. Pale boxes represent zero entries, the actual values of the nonzero entries are irrelevant for our purposes which is why they are represented by generic bold boxes. An exemplary zero column and an exemplary zero column with resolvable spoiler are show in each triangle.}
\label{fig:sierpinski}
\end{figure*}
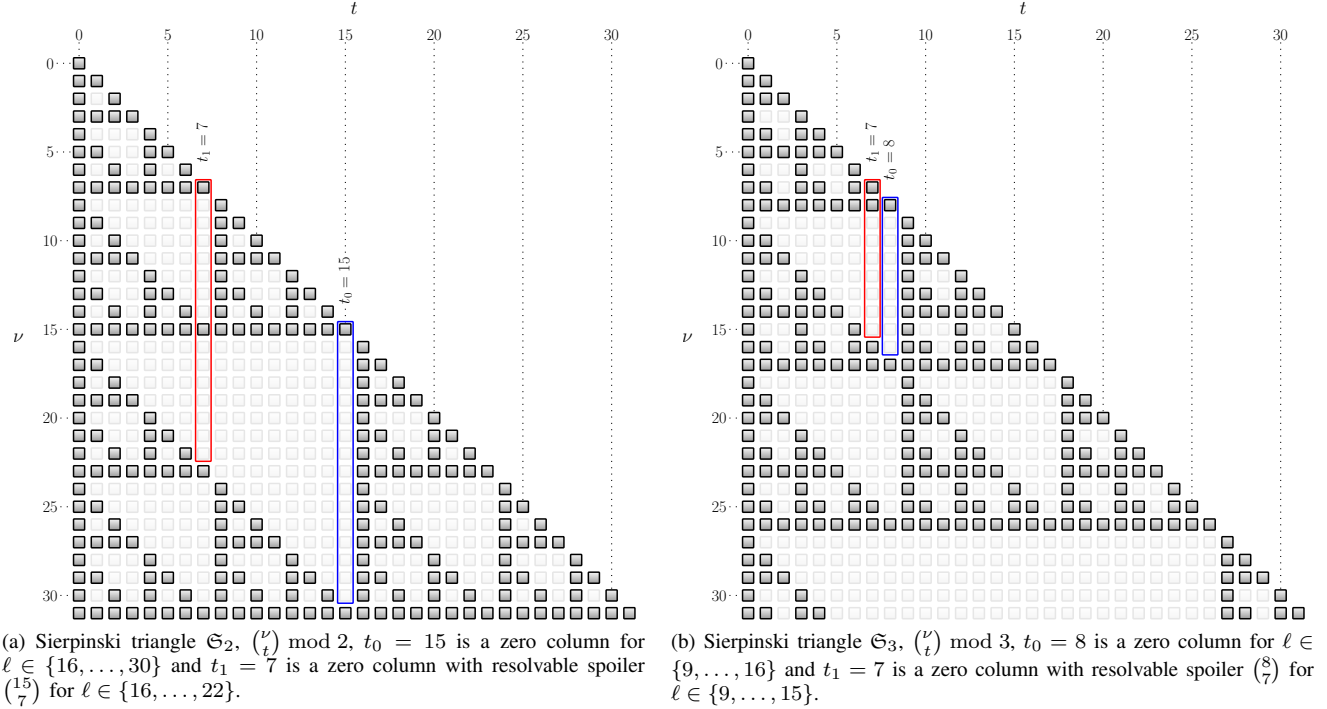

To see this, let us consider the left-aligned Pascal triangles in Fig.~\ref{fig:sierpinski}, where the entries are calculated modulo $2$ and modulo $3$, respectively. Obviously, the zero entries of the triangles follow regular patterns. Note that we use the rather uncommon left-aligned representation of the Pascal triangle in order to be able to refer to its columns. The triangle resembles variants of the left-aligned \emph{Sierpinski gasket}, one of the most basic examples of a self-similar set. The resemblance gets more accurate as more rows of the Pascal triangle are considered. In the following, we will refer to a Pascal triangle with entries modulo any positive integer $p$ as a \emph{Sierpinski triangle} and we denote it by $\mathfrak{S}_p$.

Now consider the interpolation constraints \eqref{eqn:hasse}. The summands of the outer sum are weighted by the binomial coefficients $\binom{\nu}{t}$, $\nu=t, \ldots, \ell$. These are exactly the binomial coefficients that appear at the first $\ell-t+1$ entries in column $t$ of a Sierpinski triangle. For any $\alpha\in\Fq$, $p\alpha=0$, where $p\defeq\char[\Fq]$ is the characteristic of $\Fq$. Thus, summands for which $\binom{\nu}{t}$ is a multiple of $p$ are zero.

For given list size $\ell$ and multiplicity $r$, let us assume there is a column $t_0$, $t_0<r$, in $\mathfrak{S}_p$ such that
\begin{equation}\label{eqn:zerocolumn}
  \binom{t_0+1}{t_0}, \ldots, \binom{\ell}{t_0}\equiv 0\bmod p,
\end{equation}
\ie all entries except the first one (which is $\binom{t_0}{t_0}\equiv 1\bmod p$) are zero. We refer to such columns as \emph{zero columns}\footnote{Note that $t_0<r\leq\ell$, hence the degenerate zero column with a single entry $\binom{\ell}{\ell}\equiv 1\bmod p$ cannot occur in context of the GSA. However, it will turn out to be useful for the proof of Lemma~\ref{lemma:zerocolumnsspoilers} to include this case in the definition.}. Fig.~\ref{fig:sierpinski} shows that zero columns actually exist, \eg column $t_0=15$ for $\ell\in\{16, \ldots, 30\}$ in $\mathfrak{S}_2$ (Fig.~\ref{fig:sierpinski2}) or column $t_0=8$ for $\ell\in\{9, \ldots, 16\}$ in $\mathfrak{S}_3$ (Fig.~\ref{fig:sierpinski3}).

Assume that $Q(x, z)$ is a solution of Problem~\ref{prob:interpolation} for a received vector $\vec{y}$ and consider \eqref{eqn:hasse} for zero column $t_0$. This equation simplifies to
\begin{multline}\label{eqn:sier_hasse1}
  \forall i\in\mathcal{I}\;\forall s\in\N:s<r-t_0\;\text{and}\\
  \underbrace{\binom{t_0}{t_0}z^{t_0-t_0}}_{=1}\sum_{\mu=s}^{\maxdeg{Q}{t_0}}
  \binom{\mu}{s}x^{\mu-s}Q_{t_0, \mu}%
  \bigg|_{(x, z)=\left(\alpha^{-i}, y_i\right)}=0,
\end{multline}
since all summands of the outer sum except the first one are annihilated by the zero binomial weights. But then, with the same argumentation as in the proof of Theorem~\ref{thm:reenc_factors}, the $\alpha^{-i}$, $i\in\mathcal{I}$, are roots of multiplicity $r-t_0$ of $Q_{t_0}(x)$ and thus $Q_{t_0}(x)$ can be factored as
\begin{equation}\label{eqn:factorization_zerocolumn}
  Q_{t_0}(x)=V_{t_0}(x)P_\mathcal{I}(x)^{r-t_0},
\end{equation}
where $\deg[V_{t_0}(x)]\leq \maxdeg{Q}{t_0}-n(r-t_0)$.

The following lemma specifies the conditions for the existence of a zero column $t_0$ and its location.

\begin{lemma}\label{lemma:zerocolumnexists}
Let $r,\ell\in\N\setminus\{0\}$, $r\leq\ell$. If
\begin{itemize}
  \item $\ell<p$ or
  \item $\exists\,a\in \{1,\ldots, p-1\},j\in\N:r\leq ap^j-1\leq\ell$
\end{itemize}
then a zero column does \emph{not} exist. Otherwise, find the least significant base-$p$ digit $\ell_i$ of $\ell$ such that $\ell_i<p-1$. Then,
\begin{equation*}
  t_0=\left\lfloor\frac{r}{p^{i+1}}\right\rfloor p^{i+1}-1
\end{equation*}	      
is a zero column. In particular, $t_0$ is the maximal (rightmost) zero column.
\end{lemma}

The proof of Lemma~\ref{lemma:zerocolumnexists} relies on the following well-known theorem about the divisibility of binomial coefficients, which we state in a modified form that is particularly convenient for our purposes.

\begin{theorem}[Lucas, 1878 \cite{lucas:1878}]
  Let $u,v\in\N$ and $p$ prime. Then $\binom{u}{v}\equiv 0\bmod p$ if and only if at least one base-$p$ digit of $v$ is greater than the corresponding base-$p$ digit of $u$.
\end{theorem}

\begin{IEEEproof}[Proof of Lemma~\ref{lemma:zerocolumnexists}]
A zero column $t_0$ is defined by property \eqref{eqn:zerocolumn}, which can --- using Lucas' Theorem --- be equivalently formulated as follows: $t_0$ in base $p$ must be such that it has at least one base-$p$ digit greater than that of $t_0+1$, $\ldots$, at least one base-$p$ digit greater than that of $\ell$.

If $\ell<p$ then $t_0, \ldots, \ell$ are all single-digit numbers and since $t_0<r\leq\ell$, $t_0$ cannot be greater than $t_0+1, \ldots, \ell$, proving the first case. The base-$p$ expansion of an integer that fulfills the second case is
\begin{equation*}
  (0, \ldots, 0, a, p-1, \ldots, p-1)_p,
\end{equation*}
where $a\in \{1,\ldots, p-1\}$. That is, all digits except the leading one assume the maximal possible value $p-1$, which cannot be exceeded by the digits of any integer modulo $p$, particularly not by those of $t_0$. But the most significant digit of $t_0$ cannot be larger than $a$ as well since $t_0<r\leq ap^j-1$. Thus, Lucas' Theorem cannot be fulfilled for $v=t_0$ and $u=ap^j-1$, proving the second case.

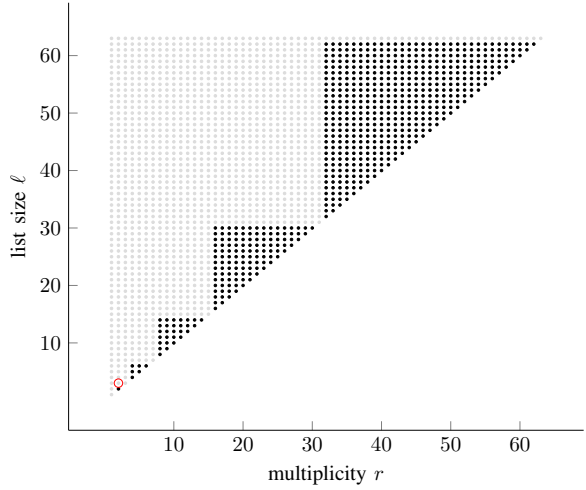
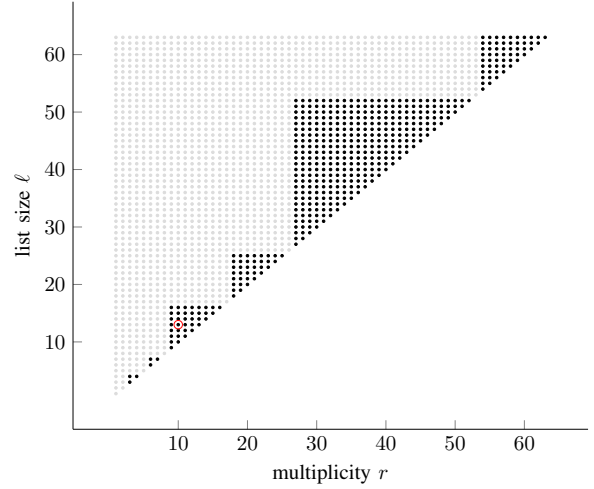
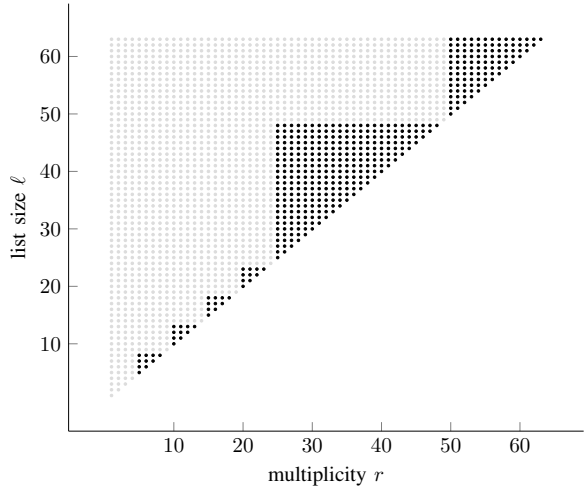
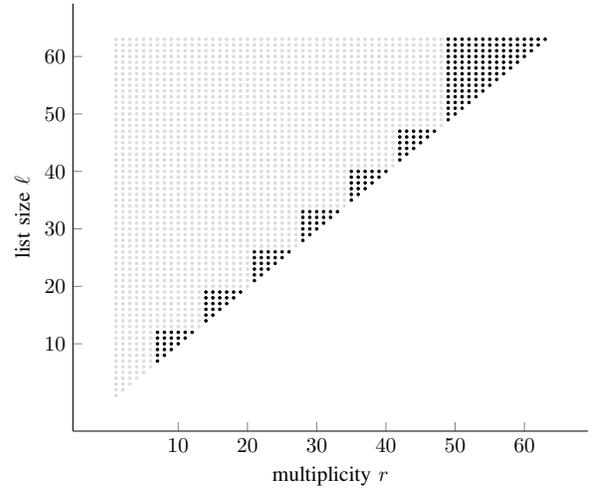
\begin{figure*}[!t]
\centering
\subfloat[$p=2$, the GSA parameters $(r, \ell)=(2, 3)$ of Examples~\ref{ex:reenc1} and \ref{ex:reenc2} are marked by a small circle. For these parameters, $t_0$ does not exist.]{
  \begin{tikzpicture}[scale=0.8,every node/.style={transform shape}]
  \end{tikzpicture}
\label{fig:zerocolumnexists2}}
\hfil
\subfloat[$p=3$, the GSA parameters $(r, \ell)=(10, 13)$ of Examples~\ref{ex:sier1} to \ref{ex:matrix1} are marked by a small circle. For these parameters, $t_0$ (besides additional zero columns with resolvable spoilers) exists.]{
  \begin{tikzpicture}[scale=0.8,every node/.style={transform shape}]
  \end{tikzpicture}
\label{fig:zerocolumnexists3}}\\
\subfloat[$p=5$]{
  \begin{tikzpicture}[scale=0.8,every node/.style={transform shape}]
  \end{tikzpicture}
\label{fig:zerocolumnexists5}}
\hfil
\subfloat[$p=7$]{
  \begin{tikzpicture}[scale=0.8,every node/.style={transform shape}]
  \end{tikzpicture}
\label{fig:zerocolumnexists7}}
\caption{Graphical representation of Lemma~\ref{lemma:zerocolumnexists}. GSA parameter combinations $(r, \ell)$ for which a zero column exists are shown as black dots, all other permissible ($r\leq\ell$) parameter combinations as light gray dots. Roughly speaking, zero columns can only exist if $r$ is close to $\ell$, which is the practically most relevant case. Note that the Sudan- and Welch--Berlekamp algorithms are represented by the dots of the leftmost column and the lower left dot, respectively.}
\label{fig:zerocolumnexists}
\end{figure*}

If the first and second cases do not apply then $\ell$ has at least two digits, at least one of them being smaller than $p-1$. This guarantees the existence of the least significant such digit, \ie $\ell_i$. We need to find the greatest $b\in\N$, $b<r$, whose digits $b_{i'}$, $i'\leq i$ are zero. This is only possible if $b$ is a multiple of $p^{i+1}$. Set $b$ to the greatest such multiple smaller than $r$, that is $b=\left\lfloor\nicefrac{r}{p^{i+1}}\right\rfloor p^{i+1}$. Let $c=b-1$. This subtraction leads to a carry up to digit $c_i$ and thus $c_{i'}=p-1$ for all $i'\leq i$, \ie the $i$ least significant digits of $c$ are maximal. Due to the maximality of $b$ there can be no $d=ap^{i+1}-1$ for which both $d_{i'}=p-1$ for all $i'\leq i$ and $c<d<r$ hold. By invalidity of the second case there can also be no such $d$ with $r\leq d\leq\ell$. This shows that Lucas' Theorem is fulfilled for $v=c$ and all $u=c+1, \ldots, \ell$, thereby proving the third case. The maximality of $t_0$ follows from the maximality of $b$.
\end{IEEEproof}

\begin{example}\label{ex:sier1}
Consider the conventional RS code $\GRS_{\mathcal{A}, \mathcal{B}}(\F_{27}; 26, 16)$ with $\mathcal{A}=\{\alpha^i:i=0, \ldots, 25\}$ and $\mathcal{B}=\{1, \ldots, 1\}$. The characteristic of the code's base field is $\char[\F_{27}]=3$. For this code, the GSA with multiplicity $r=10$ and list size $\ell=13$ can correct up to $\varepsilon_0=6$ errors. Lemma~\ref{lemma:zerocolumnexists} yields $t_0=\left\lfloor\nicefrac{11}{3}\right\rfloor3-1=8$, since the binary expansion of $\ell=13$ is $(1, 1, 1)_3$. We can easily check Fig.~\ref{fig:sierpinski3} in order to confirm that this is a zero column for $\ell=13$. Thus, according to \eqref{eqn:factorization_zerocolumn}, for any result
\begin{equation*}
  Q(x, z)=Q_0(x)+Q_1(x)z+\cdots+Q_{8}(x)z^8+\cdots+Q_{13}(x)z^{13}
\end{equation*}
of the interpolation step of the GSA holds the factorization $Q_8(x)=V_8(x)P_\mathcal{I}(x)^2$, where $P_\mathcal{I}(x)=x^{26}+2$ (from \eqref{eqn:Px}) and $\maxdeg{V}{8}=27$ (from $\maxdeg{Q}{8}=79$).
\end{example}

It follows directly from Lemma~\ref{lemma:zerocolumnexists} that zero columns cannot exist for the special cases $r=1<\ell$ (Sudan algorithm) and $r=1=\ell$ (Welch--Berlekamp algorithm) of the GSA. The statement of the lemma can be interpreted graphically, see Fig.~\ref{fig:zerocolumnexists}.

In the following, we will generalize the concept of zero columns to zero columns with resolvable spoilers. This will reveal additional structure in solutions $Q(x, z)$ of Problem~\ref{prob:interpolation}, \ie to factorizations of additional univariate polynomials $Q_\nu(x)$, $\nu\in\{0, \ldots, r-1\}$, besides $Q_{t_0}(x)$.

Assume that for given GSA parameters $r, \ell$ and base field $\Fq$ with $p=\char[\Fq]$ there exists a zero column $t_0$ in $\mathfrak{S}_p$. Further assume there is a column $t_1$, $t_1<t_0$, in  $\mathfrak{S}_p$ such that
\begin{multline*}
  \binom{t_0}{t_1}\not\equiv 0\bmod p\;\text{and}\\\forall \nu=t_1+1, \ldots, \ell, \nu\neq t_0:\binom{\nu}{t_1}\equiv 0\bmod p.
\end{multline*}
We refer to such a column as \emph{zero column with spoiler at $\binom{t_0}{t_1}$}, examples are shown in Fig.~\ref{fig:sierpinski2} and Fig.~\ref{fig:sierpinski3}, respectively.

If $Q(x, z)$ is a solution of Problem~\ref{prob:interpolation} for a received vector $\vec{y}$ then \eqref{eqn:hasse} for column $t_1$, $t_1<t_0$, becomes
\begin{multline}\label{eqn:sier_hasse2}
  \forall i\in\mathcal{I}\;\forall s\in\N:s<r-t_0\;\text{and}\\
	  \underbrace{\binom{t_1}{t_1}z^{t_1-t_1}}_{=1}\sum_{\mu=s}^{\maxdeg{Q}{t_1}}
	  \binom{\mu}{s}x^{\mu-s}Q_{t_1, \mu}+\\
	  \binom{t_0}{t_1}z^{t_0-t_1}\sum_{\mu=s}^{\maxdeg{Q}{t_0}}
	  \binom{\mu}{s}x^{\mu-s}Q_{t_0, \mu}
	  \bigg|_{(x, z)=\left(\alpha^{-i}, y_i\right)}=0,
\end{multline}
since all summands except the ones for $\nu=t_1$ and $\nu=t_0$ are annihilated. But, according to \eqref{eqn:sier_hasse1}, the second sum evaluates to zero at all $\alpha^{-i}$, $i\in\mathcal{I}$, since $t_0$ is by assumption a zero column. We refer to $\binom{t_0}{t_1}$ as a \emph{resolvable spoiler} for $t_1$ because the sum associated with $\binom{t_0}{t_1}$ vanishes. As a result, we obtain
\begin{multline*}
  \forall i\in\mathcal{I}\;\forall s\in\N:s<r-t_0\;\text{and}\\
	  \sum_{\mu=s}^{\maxdeg{Q}{t_1}}\binom{\mu}{s}x^{\mu-s}Q_{t_1, \mu}
	  \bigg|_{x=\alpha^{-i}}=0,
\end{multline*}
\ie the $\alpha^{-i}$, $i\in\mathcal{I}$, are roots of multiplicity $r-t_0$ of $Q_{t_1}(x)$ and thus it can be factored as
\begin{equation*}
  Q_{t_1}(x)=V_{t_1}(x)P_\mathcal{I}(x)^{r-t_0},
\end{equation*}
where $\deg[V_{t_1}(x)]\leq \maxdeg{Q}{t_1}-n(r-t_0)$.

It is easy to see that if $\binom{t_1}{t_2}$ is a spoiler for $t_2$, $t_2<t_1$, then it is also resolvable. Furthermore, it is easy to see that the concept generalizes to multiple spoilers. This leads to the following recursive definition:

\begin{definition}
Consider $\mathfrak{S}_p$ and $r,\ell\in\N\setminus\{0\}$. For $\nu\leq t_0$ let
\begin{multline*}   
  \mathcal{R}^{(\ell)}_\nu\defeq\{t\in\N:\nu<t\leq t_0,\\t\;\text{is a zero column with resolvable spoilers}\}
\end{multline*}
and
\begin{equation*}  
  \mathcal{S}^{(\ell)}_\nu\defeq\left\{t\in\N:\binom{t}{\nu}\;\text{is a spoiler for}\;\nu\right\}.
\end{equation*}
Then $\nu$ is a \emph{zero column with resolvable spoilers} if and only if $\mathcal{S}^{(\ell)}_\nu\subseteq \mathcal{R}^{(\ell)}_\nu$. The basic case is $\mathcal{R}^{(\ell)}_{t_0}=\{t_0\}$ if $t_0$ exists.
\end{definition}

Note that zero columns are special cases of zero columns with resolvable spoilers where $\mathcal{S}^{(\ell)}_\nu=\emptyset$. The sets of zero columns with resolvable spoilers are non-increasing with $\nu$, \ie $\mathcal{R}_{\nu+1}^{(\ell)}\subseteq \mathcal{R}_\nu^{(\ell)}$. We stress that $\mathcal{R}^{(\ell)}_0$ contains all zero columns with resolvable spoilers in $\mathfrak{S}_p$, \ie it is the set that we are interested in.

\begin{lemma}\label{lemma:zerocolumnsspoilers}
Let $p=\char[\Fq]$ and $r,\ell\in\N\setminus\{0\}$, $r\leq\ell$. If according to Lemma~\ref{lemma:zerocolumnexists} a maximal zero column $t_0$ of $\mathfrak{S}_p$ exists then the set of zero columns with resolvable spoilers of $\mathfrak{S}_p$ is
\begin{equation*}
  \mathcal{R}^{(\ell)}_0=\left\{t\in\N: t<r\;\text{and}\; \sum_{\ell'=t_0+1}^{\ell}\left(\binom{\ell'}{t}\bmod p\right)=0\right\},
\end{equation*}
otherwise it is $\mathcal{R}^{(\ell)}_0=\emptyset$.
\end{lemma}

\begin{IEEEproof}
Assume that $t_0$ exists. Note that it is the maximal zero column not only for $\ell$ but for all $\ell'=t_0, \ldots, \ell$ since appending an entry to a nonzero column obviously cannot make it a zero column. Let us set $\ell'=t_0+1$. Note that in that case, column $t_0$ is degenerate. We obtain $\mathcal{R}^{(t_0+1)}_{t_0-1}=\{t_0\}$. If we now advance to column $t_0-1$ it can either have a spoiler at $\binom{t_0}{t_0+1}$ or no spoiler at all, meaning either $\mathcal{S}^{(t_0+1)}_{t_0-1}=\{t_0\}$ or $\mathcal{S}^{(t_0+1)}_{t_0-1}=\emptyset$. In both cases $\mathcal{S}^{(t_0+1)}_{t_0-1}\subseteq\mathcal{R}^{(t_0+1)}_{t_0-1}$ and we have $\mathcal{R}^{(t_0+1)}_{t_0-2}=\{t_0-1, t_0\}$. In the same manner, column $t_0-2$ can either have $\mathcal{S}^{(t_0+1)}_{t_0-2}=\{t_0-1\}$, $\mathcal{S}^{(t_0+1)}_{t_0-2}=\{t_0\}$, $\mathcal{S}^{(t_0+1)}_{t_0-2}=\{t_0-1, t_0\}$, or $\mathcal{S}^{(t_0+1)}_{t_0-2}=\emptyset$. In all cases $\mathcal{S}^{(t_0+1)}_{t_0-2}\subseteq\mathcal{R}^{(t_0+1)}_{t_0-2}$, resulting in $\mathcal{R}^{(t_0+1)}_{t_0-3}=\{t_0-2, t_0-1, t_0\}$. This process continues all the way to column $0$ and as a result \emph{all} spoilers are resolvable and thus \emph{all} columns of $\mathfrak{S}_p$ are zero columns with resolvable spoilers with respect to $\ell'=t_0+1$, eventually resulting in $\mathcal{R}^{(t_0+1)}_{0}=\{0, \ldots, t_0\}$.

Now let us gradually increase $\ell'$ by one until $\ell'=\ell$ and recall that $t_0$ stays a maximum zero column. Due to the maximality of $t_0$, any nonzero $\binom{\ell'}{t}\not\equiv 0\bmod p$ with $0\leq t\leq t_0$ is a non-resolvable spoiler and thus such columns $t$ must be removed from $\mathcal{R}^{(\ell')}_{0}$. But this means that eventually only the columns $t$ for which all $\binom{\ell'}{t}$, $\ell'=t_0+1, \ldots, \ell$, are zero remain in $\mathcal{R}^{(\ell)}_{0}$, which (after substituting $t$ by $\nu$) proves the statement.

If $t_0$ does not exist then there exists no zero column and also no resolvable spoilers, hence $\mathcal{R}^{(\ell)}_{0}=\emptyset$.
\end{IEEEproof}

\begin{example}\label{ex:sier2}
Lemma~\ref{lemma:zerocolumnexists} yields maximal zero column $t_0=8$ for the setting of Example~\ref{ex:sier1}. Fig.~\ref{fig:sierpinski3} shows that for $t_1=7$ we have $\mathcal{S}^{(13)}_7=\{8\}=\mathcal{R}^{(13)}_7$, \ie $t_1=7$ is a zero column with resolvable spoilers and we can set $\mathcal{R}^{(13)}_6=\{7, 8\}$. For $t_2=6$ we have $\mathcal{S}^{(13)}_6=\{7, 8\}=\mathcal{R}^{(13)}_6$ and thus it is a zero column with resolvable spoilers as well. This gives $\mathcal{R}^{(13)}_5=\{6, 7, 8\}$. For $t_3=5$ we have $\mathcal{S}^{(13)}_5=\{8\}\subseteq\mathcal{R}^{(13)}_5$ and thus it is a zero column with resolvable spoilers as well. It turns out that for all $t_4<t_3$ holds $\mathcal{S}^{(13)}_{t_4}\not\subseteq\{5, 6, 7, 8\}=\mathcal{R}^{(13)}_{t_4}$ and thus the only zero columns with resolvable spoilers in $\mathfrak{S}_3$ with respect to $\ell=13$ are $t_0=8$, $t_1=7$, $t_2=6$, and $t_3=5$. It can be readily checked that Lemma~\ref{lemma:zerocolumnsspoilers} confirms this result and delivers $\mathcal{R}^{(13)}_0=\{5, 6, 7, 8\}$.
\end{example}

The following map will turn out to be useful in the following, it returns either $\nu$ itself or its greatest spoiler:
\begin{equation}\label{eqn:gnu}
  g ~:~ \left\{\begin{array}{rcl} \N &\to &\N\\
   \nu &\mapsto &\left\{\begin{array}[]{ll}
                 \max\{\mathcal{S}_\nu^{(\ell)}\} & \mathcal{S}_\nu^{(\ell)}\neq\emptyset\\
                 \nu & \mathcal{S}_\nu^{(\ell)}=\emptyset\\
               \end{array}\right.
   \end{array}\right..
\end{equation}

\begin{theorem}\label{thm:sier_factors}
Let $\GRSparloc$ be a GRS code and $r,\ell$ such that the GSA can correct at most $\varepsilon_0$ errors. Let further $\vec{c}\in\GRS$, $\vec{e}\in\Fq^n$ with $\Hw{\vec{e}}\leq\varepsilon_0$ and $\vec{y}=\vec{c}+\vec{e}$. When the GSA is applied to $\vec{y}$ it yields a bivariate result polynomial $Q(x, z)=Q_0(x)+Q_1(x)z+\cdots +Q_\ell(x)z^\ell\in\Fq[x, z]$ whose constituent univariate polynomials $Q_\nu(x)$, $\nu\in\mathcal{R}_0^{(\ell)}$, can be factored as
\begin{equation}\label{eqn:sier_factors}
  Q_\nu(x)=V_\nu(x)P_\mathcal{I}(x)^{r-g[\nu]},
\end{equation}
where $\deg[V_\nu(x)]\leq \maxdeg{Q}{\nu}-n\left(r-g[\nu]\right)\defeq\maxdeg{V}{\nu}$.
\end{theorem}

\begin{IEEEproof}
Let $\nu\in\mathcal{R}_0^{(\ell)}$. Since $\nu$ is a zero column with resolvable spoilers, all spoilers $\binom{t}{\nu}$ with $t\in\mathcal{S}_\nu^{(\ell)}$ are resolvable, that is,
\begin{multline}\label{eqn:resolvablespoilers}
  \forall t\in\mathcal{S}_\nu^{(\ell)}\forall i\in\mathcal{I}\;\forall s\in\N:s<r-t\;\text{and}\\
  \sum_{\mu=s}^{\maxdeg{Q}{t}}\binom{\mu}{s}x^{\mu-s}Q_{t, \mu}\bigg|_{x=\alpha^{-i}}=0.
\end{multline}
Since by definition all terms except the ones weighted by the spoilers vanish, we can write \eqref{eqn:hasse} as\footnote{Note that this is the generalization of \eqref{eqn:sier_hasse2} to multiple spoilers.}
\begin{multline*}
  \forall i\in\mathcal{I}\;\forall s\in\N:s<r-\nu\;\text{and}\\
	  \sum_{t\in\mathcal{S}_\nu^{(\ell)}}\binom{t}{\nu}z^{t-\nu}\sum_{\mu=s}^{d_{t}}
	  \binom{\mu}{s}x^{\mu-s}Q_{t, \mu}+\\
	  \sum_{\mu=s}^{\maxdeg{Q}{\nu}}
	  \binom{\mu}{s}x^{\mu-s}Q_{\nu, \mu}
	  \bigg|_{(x, z)=\left(\alpha^{-i}, y_i\right)}=0.
\end{multline*}
In order to let the sum over $t$ vanish in the case $\mathcal{S}_\nu^{(\ell)}\neq\emptyset$ (\ie to exploit \eqref{eqn:resolvablespoilers}), we must guarantee $s<r-t$ for all $t\in\mathcal{S}_\nu^{(\ell)}$, \ie $s<r-\max\{\mathcal{S}_\nu^{(\ell)}\}$. In case $\mathcal{S}_\nu^{(\ell)}=\emptyset$ the sum over $t$ is empty and thus it is sufficient to guarantee $s<r-\nu$. Due to the definition \eqref{eqn:gnu} of $g[\nu]$ we have $s<r-g[\nu]$ in both cases and thus
\begin{multline*}
  \forall i\in\mathcal{I}\;\forall s\in\N:s<r-g_\nu\;\text{and}\\
  \sum_{\mu=s}^{\maxdeg{Q}{\nu}}
	  \binom{\mu}{s}x^{\mu-s}Q_{\nu, \mu}
	  \bigg|_{x=\alpha^{-i}}=0
\end{multline*}
and the $\alpha^{-i}$, $i\in\mathcal{I}$, are roots of multiplicity $r-g[\nu]$ of $Q_\nu(x)$ and thus it can be factored as in \eqref{eqn:sier_factors}. The bound on the degrees of the $V_\nu(x)$ follows from a comparison of the involved polynomial degrees.
\end{IEEEproof}

Just as the re-encoding prefactors $P_\mathcal{J}(x)^{r-\nu}$, $\nu=0, \ldots, r-1$, from Section~\ref{sec:reenc}, the \emph{Sierpinski prefactors} $P_\mathcal{I}(x)^{r-g[\nu]}$, $\nu\in\mathcal{R}_0^{(\ell)}$, are fixed a-priori and do not depend on the received vector $\vec{y}$.

\begin{example}\label{ex:sier3}
We have already seen with the help of Lemma~\ref{lemma:zerocolumnexists} that for the setting of Example~\ref{ex:sier1} any result of the polynomial step of the GSA contains a univariate constituent polynomial $Q_8(x)$ with factorization $Q_8(x)=V_8(x)P_\mathcal{I}(x)^2$, where $P_\mathcal{I}(x)=x^{26}+2$ and $\maxdeg{V}{8}=27$. From Example~\ref{ex:sier2} we have $\mathcal{R}_0^{(13)}=\{5, 6, 7, 8\}$, \ie Theorem~\ref{thm:sier_factors} additionally guarantees factorizations of $Q_5(x)$, $Q_6(x)$, and $Q_7(x)$. They are
\begin{align*}
  Q_5(x) &= V_5(x)P_\mathcal{I}(x)^{2}\\
  Q_6(x) &= V_6(x)P_\mathcal{I}(x)^{2}\\
  Q_7(x) &= V_7(x)P_\mathcal{I}(x)^{2},
\end{align*}
with $\maxdeg{V}{5}=72$, $\maxdeg{V}{6}=57$, and $\maxdeg{V}{7}=42$, respectively, because $g[5]=g[6]=g[7]=8$, $\maxdeg{Q}{5}=124$, $\maxdeg{Q}{6}=109$, and $\maxdeg{Q}{7}=94$.

The situation is depicted in Fig.~\ref{fig:sier_ex}. For the zero columns with resolvable spoilers $t_0=8$, $t_1=7$, $t_2=6$, and $t_3=5$ the summands $\binom{\ell'}{t}$ from Lemma~\ref{lemma:zerocolumnsspoilers} are enclosed by dashed rectangles and the sets of (resolvable) spoilers $\mathcal{S}_{\nu}^{(13)}$, $\nu=5, \ldots, 7$, are enclosed by solid rectangles. The greatest spoilers $g[\nu]$ are the lowermost entries within the solid rectangles.

\begin{figure}[!t]
\centering
  \begin{tikzpicture}[scale=\pascalscale,every node/.style={transform shape}]    
  \end{tikzpicture}
\caption{Visualization of Examples~\ref{ex:sier2} and \ref{ex:sier3}.}
\label{fig:sier_ex}
\end{figure}
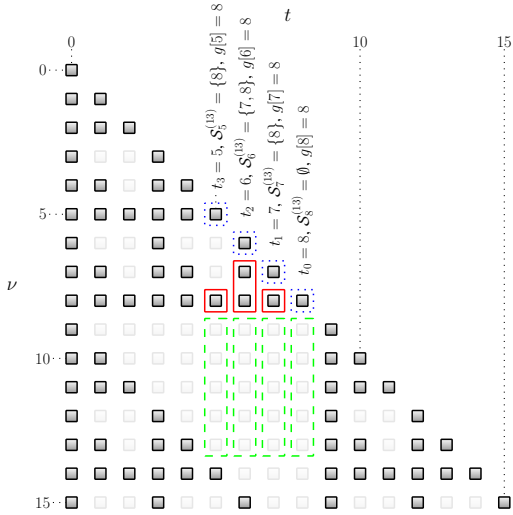

\end{example}

\section{Exploiting Prefactors}\label{sec:factors}

We have seen in Sections~\ref{sec:reenc} and \ref{sec:sierpinski} that some of the univariate constituent polynomials of the GSA interpolation step (Problem~\ref{prob:interpolation}) have certain prefactors that are fixed a-priori. We will show in this section how this knowledge can be exploited in order to simplify solving the interpolation step. More precisely, we show that the associated linear system of equations in $\sum_{\nu=0}^\ell (\maxdeg{Q}{\nu}+1)$ unknowns can be reduced to a linear system of smaller size.

So far, we have treated re-encoding and Sierpinski prefactors separately. This is not practical, one obviously wishes to exploit both types of prefactors jointly whenever possible. Since re-encoding and Sierpinski prefactors of a given univariate constituent polynomial $Q_\nu(x)$, $\nu\in\{0, \ldots, r-1\}$ are in general not coprime, we have to calculate the \emph{combined prefactors} in the following manner.

\begin{theorem}\label{thm:combined_factors}
Let $\GRSparloc$ be a GRS code, $\mathcal{J}\subseteq\mathcal{I}$ with $\vert\mathcal{J}\vert=k$, and $r,\ell$ such that the GSA can correct at most $\varepsilon_0$ errors. Let further $\vec{c}\in\GRS$, $\vec{e}\in\Fq^n$ with $\Hw{\vec{e}}\leq\varepsilon_0$ and $\vec{y}=\vec{c}+\vec{e}$. When the GSA is applied to $\REENCf{\mathcal{J}}[\vec{y}]$ it yields a bivariate result polynomial $Q(x, z)=Q_0(x)+Q_1(x)z+\cdots +Q_\ell(x)z^\ell\in\Fq[x, z]$ whose univariate constituent polynomials $Q_\nu(x)$ can be factored as
\begin{equation*}
    Q_\nu(x)=\left\{\begin{array}[]{ll}
                 W_\nu(x)P_\mathcal{J}(x)^{r-\nu}P_{\mathcal{I}\setminus\mathcal{J}}(x)^{r-g[\nu]} & \nu<r, \nu\in\mathcal{R}_0^{(\ell)}\\
                 U_\nu(x)P_\mathcal{J}(x)^{r-\nu} & \nu<r,\nu\not\in\mathcal{R}_0^{(\ell)}
               \end{array}\right..
\end{equation*}The degrees of the $W_\nu(x)$ and $U_\nu(x)$ are bounded by
\begin{equation*}
  \deg[W_\nu(x)]\leq \maxdeg{Q}{\nu}-k(r-\nu)-(n-k)(r-g[\nu])\defeq\maxdeg{W}{\nu}
\end{equation*}
and $\deg[U_\nu(x)]\leq \maxdeg{U}{\nu}$, respectively.
\end{theorem}

\begin{IEEEproof}
The second case is a repetition of Theorem~\ref{thm:reenc_factors} and there is nothing left to prove. As for the first case, we have $\mathcal{J}\subseteq\mathcal{I}$. This allows to write the Sierpinski prefactors from Theorem~\ref{thm:sier_factors} as $P_\mathcal{I}(x)^{r-g[\nu]}=P_\mathcal{J}(x)^{r-g[\nu]}P_{\mathcal{I}\setminus\mathcal{J}}(x)^{r-g[\nu]}$. Since $\binom{\nu}{g[\nu]}$ is a spoiler, we have $g[\nu]\leq \nu$. This allows to write the re-encoding prefactors from Theorem~\ref{thm:reenc_factors} as $P_\mathcal{J}^{r-\nu}=P_\mathcal{J}(x)^{r-g[\nu]}P_\mathcal{J}(x)^{\nu-g[\nu]}$. The factor $P_\mathcal{J}(x)^{r-g[\nu]}$ is common to both factorizations (Sierpinski and re-encoding) and must be counted not more than once in a combined factorization. Hence, we obtain
\begin{equation*}
  Q_\nu(x)=W_\nu(x)\underbrace{P_\mathcal{J}(x)^{r-g[\nu]}P_\mathcal{J}(x)^{\nu-g[\nu]}}_{=P_\mathcal{J}^{r-\nu}}P_{\mathcal{I}\setminus\mathcal{J}}(x)^{r-g[\nu]},
\end{equation*}
which proves the first case. The degree constraints follow directly from the degree constraints in Theorems~\ref{thm:reenc_factors} and \ref{thm:sier_factors}.
\end{IEEEproof}

In the following, it will be convenient to have the sets
\begin{align*}
  \mathcal{F}&\defeq\{\nu:0\leq\nu\leq\ell,Q_\nu(x)\;\text{has prefactor}\}\shortintertext{and}
  \mathcal{F}^c&\defeq\{\nu:0\leq\nu\leq\ell,\nu\notin\mathcal{F}\}.
\end{align*}

\begin{example}\label{ex:sier4}
  For the setting of Example~\ref{ex:sier1}, a solution of the GSA interpolation step Problem~\ref{prob:interpolation} for the projected received vector $\REENCf{\mathcal{J}}[\vec{y}]$ with $\mathcal{J}=\{10, \ldots, 25\}$ could be
\begin{equation*}
  Q(x, z)=Q_0(x)+Q_1(x)z+\cdots+Q_{13}(x)z^{13}.
\end{equation*}
The re-encoding prefactors for $\nu=0, \ldots, 9$ according to Theorem~\ref{thm:reenc_factors} are $P_\mathcal{J}(x)^{10-\nu}$, where
\begin{multline*}
  P_\mathcal{J}(x)=x^{16} + \alpha^{25}z^{15} + \alpha^{10}z^{14} + \alpha^{13}z^{13} + \alpha^{23}z^{12}\\
  + \alpha^{16}z^{11} + \alpha^5z^{10} + \alpha^{18}z^9 + \alpha^{15}z^8 + \alpha^9z^7 + \alpha^{13}z^6\\
  + \alpha^{15}z^5 + \alpha^{13}z^4 + \alpha^{20}z^3 + \alpha^8z^2 + \alpha^{14}z + \alpha^6.
\end{multline*}
We have
\begin{multline*}
  P_{\mathcal{I}\setminus\mathcal{J}}=z^{10} + \alpha^{12}z^9 + \alpha^6z^8 + \alpha^2z^7 + \alpha^{20}z^6\\
  + \alpha^{18}z^5 + \alpha^{11}z^4 + \alpha^{10}z^3 + \alpha^5z^2 + \alpha^2z + \alpha^7.
\end{multline*}
and from Example~\ref{ex:sier2} we have $\mathcal{R}_7^{(13)}=\{8\}$, $\mathcal{R}_6^{(13)}=\{7, 8\}$, $\mathcal{R}_5^{(13)}=\{6, 7, 8\}$, and, finally,  $\mathcal{R}_0^{(13)}=\{5, 6, 7, 8\}$. $\mathcal{R}_8^{(13)}=\{8\}$ since $t_0=8$ is a zero column. With $g[5]=g[6]=g[7]=g[8]=8$ we obtain the combined prefactors $P_\mathcal{J}(x)^{10-5}P_{\mathcal{I}\setminus\mathcal{J}}(x)^{10-8}$, $P_\mathcal{J}(x)^{10-6}P_{\mathcal{I}\setminus\mathcal{J}}(x)^{10-8}$, $P_\mathcal{J}(x)^{10-7}P_{\mathcal{I}\setminus\mathcal{J}}(x)^{10-8}$, and $P_\mathcal{J}(x)^{10-8}P_{\mathcal{I}\setminus\mathcal{J}}(x)^{10-8}$ of $Q_5(x)$, $Q_6(x)$, $Q_7(x)$, and $Q_8(x)$, respectively. Re-encoding-, Sierpinski- and combined prefactors for the code under consideration are shown in Table~\ref{tab:sier_example}. The table contains upper bounds on the degrees of the quotient polynomials (after dividing by the prefactors) for each of the factorizations. Re-encoding and combined prefactors for this example give $\mathcal{F}=\{0, \ldots, 9\}$, while Sierpinski prefactors give $\mathcal{F}=\{5, 6, 7, 8\}$.

\begin{table*}[!t]
\renewcommand{\arraystretch}{1.3}
\caption{Re-encoding-, Sierpinski- and combined factorizations of the univariate constituent polynomials of bivariate GSA interpolation polynomials $Q(x, z)$ for the setting of Examples~\ref{ex:sier1}--\ref{ex:sier4}, \ie $\GRS_{\mathcal{A}, \mathcal{B}}(\F_{27}; 26, 16)$ with $\mathcal{A}=\{\alpha^i:i=0, \ldots, 25\}$, $\mathcal{B}=\{1, \ldots, 1\}$, and GSA parameters $(r, \ell)=(10, 13)$.}
\label{tab:sier_example}
\centering
\newcolumntype{C}[1]{>{\centering\let\newline\\\arraybackslash\hspace{0pt}}m{#1}}
\begin{tabular}{|C{0.4cm}||C{1.5cm}||C{2cm}|C{1.5cm}||C{2cm}|C{1.5cm}||C{3.3cm}|C{1.5cm}|}
\hline
$\nu$ & bound $\maxdeg{Q}{\nu}$ & re-encoding factorization & bound $\maxdeg{U}{\nu}$ & Sierpinski factorization & bound $\maxdeg{V}{\nu}$ & combined factorization & bound $\maxdeg{U}{\nu}$, $\maxdeg{W}{\nu}$\\
\hline\hline
0      &  199 & $U_0(x)P_\mathcal{J}(x)^{10}$ &  39 & ---                        &  199 & $U_0(x)P_\mathcal{J}(x)^{10}$ &  39\\
1      &  184 & $U_1(x)P_\mathcal{J}(x)^9$    &  40 & ---                        &  184 & $U_1(x)P_\mathcal{J}(x)^9$    &  40\\
2      &  169 & $U_2(x)P_\mathcal{J}(x)^8$    &  41 & ---                        &  169 & $U_2(x)P_\mathcal{J}(x)^8$    &  41\\
3      &  154 & $U_3(x)P_\mathcal{J}(x)^7$    &  42 & ---                        &  154 & $U_3(x)P_\mathcal{J}(x)^7$    &  42\\
4      &  139 & $U_4(x)P_\mathcal{J}(x)^6$    &  43 & ---                        &  139 & $U_4(x)P_\mathcal{J}(x)^6$    &  43\\
5      &  124 & $U_5(x)P_\mathcal{J}(x)^5$    &  44 & $V_5(x)P_\mathcal{I}(x)^2$ &   72 & $W_5(x)P_\mathcal{J}(x)^5 P_{\mathcal{I}\setminus\mathcal{J}}(x)^2$ &  24\\
6      &  109 & $U_6(x)P_\mathcal{J}(x)^4$    &  45 & $V_6(x)P_\mathcal{I}(x)^2$ &   57 & $W_6(x)P_\mathcal{J}(x)^4 P_{\mathcal{I}\setminus\mathcal{J}}(x)^2$ &  25\\
7      &   94 & $U_7(x)P_\mathcal{J}(x)^3$    &  46 & $V_7(x)P_\mathcal{I}(x)^2$ &   42 & $W_7(x)P_\mathcal{J}(x)^3 P_{\mathcal{I}\setminus\mathcal{J}}(x)^2$ &   26\\
8      &   79 & $U_8(x)P_\mathcal{J}(x)^2$    &  47 & $V_8(x)P_\mathcal{I}(x)^2$ &   27 & $W_8(x)P_\mathcal{J}(x)^2 P_{\mathcal{I}\setminus\mathcal{J}}(x)^2$ &   27\\
9      &   64 & $U_9(x)P_\mathcal{J}(x)$      &  48 & ---                        &   64 & $U_9(x)P_\mathcal{J}(x)$       &  48\\
10     &   49 & ---                           &  49 & ---                        &   49 & ---                            &  49\\
11     &   34 & ---                           &  34 & ---                        &   34 & ---                            &  34\\
12     &   19 & ---                           &  19 & ---                        &   19 & ---                            &  19\\
13     &    4 & ---                           &   4 & ---                        &    4 & ---                            &   4\\
\hline\hline
$\sum$ & 1421 &                               & 541 &                            & 1213 &                                & 461\\
\hline
\end{tabular}
\end{table*}

\end{example}

As noted before, GSA interpolation (Problem~\ref{prob:interpolation}) amounts to finding the solution of a linear system of equations. This can be done naively using Gaussian elimination. Several faster methods \cite{olshevsky_shokrollahi:2003, zeh_gentner_augot:2011} have been developed, all of which exploit the \emph{structure of the involved coefficient matrix}. We conjecture that all these methods can be applied to a reduced linear system of equations, which can be obtained using re-encoding-, Sierpinski-, or combined prefactors. The key idea here is to exploit the a-priori known \emph{structure of the solutions}, which follows from the a-priori known prefactors. This can be accomplished using the following lemma.

\begin{lemma}\label{lemma:lincomb_solution}
  For an arbitrary field $\F$ let
  \begin{equation*}
    \mat{A}=\begin{pmatrix}
      \vec{a}_0^T & \cdots & \vec{a}_{m-1}^T\\
    \end{pmatrix}\in\F^{n\times m}
  \end{equation*}
  and $\vec{b}=(b_0, \ldots, b_{n-1})^T$ be coefficient matrix and vector of constant terms of a linear system of equations $\mat{A}\vec{x}=\vec{b}$ that has at least one solution $\vec{x}=(x_0, \ldots, x_{m-1})^T$. Let $y$ be a linear combination of the solution variables $x_0, \ldots, x_{m-1}$, \ie
  \begin{equation}\label{eqn:lincomb}
    y=\sum_{i=0}^{m-1}\beta_i x_i,
  \end{equation}  
  where $\beta_0, \ldots, \beta_{m-1}\in\F$, $\beta_0\neq 0$. Then
   \begin{equation*}
    \widetilde{\mat{A}}=\begin{pmatrix}
      \vec{a}_1^T-\frac{\beta_1}{\beta_0}\vec{a}_0^T & \cdots & \vec{a}_{m-1}^T-\frac{\beta_{m-1}}{\beta_0}\vec{a}_0^T & \frac{1}{\beta_0}\vec{a}_0^T\\
    \end{pmatrix}
  \end{equation*}
  is the coefficient matrix of a linear system $\widetilde{\mat{A}}\widetilde{\vec{x}}=\vec{b}$ that has a solution $\widetilde{\vec{x}}=(x_1, \ldots, x_{m-1}, y)^T$.
\end{lemma}

\begin{IEEEproof}
  We can augment the linear system $\mat{A}\vec{x}=\vec{b}$ in order to obtain a linear system $\mat{A}'\vec{x}'=\vec{b}'$ with
   \begin{equation*}
    \mat{A}'=\begin{pmatrix}
      \vec{a}_0^T & \vec{a}_1^T & \cdots & \vec{a}_{m-1}^T & \vec{0}^T\\
      -\beta_0  & -\beta_1  & \cdots & -\beta_{m-1}  & 1\\
    \end{pmatrix}
  \end{equation*}
  and $\vec{b}'=(b_0, \ldots, b_{n-1}, 0)^T$ that has a solution $\vec{x}'=(x_0, \ldots, x_{m-1}, y)^T$. The last row of this system can be used to annihilate the column vector $\vec{a}_0$, which results in
  \begin{equation*}
    \mat{A}''=\begin{pmatrix}
      \vec{0}^T  & \vec{a}_1^T-\frac{\beta_1}{\beta_0}\vec{a}_0^T & \cdots & \vec{a}_{m-1}^T-\frac{\beta_{m-1}}{\beta_0}\vec{a}_0^T & \frac{1}{\beta_0}\vec{a}_0^T\\
      -\beta_0 & -\beta_1                                   & \cdots & -\beta_{m-1}                                       & 1\\
    \end{pmatrix}.
  \end{equation*}
  This allows to split the system into linear combination \eqref{eqn:lincomb} and a part $\widetilde{\mat{A}}\widetilde{\vec{x}}=\vec{b}$ that is independent of $x_0$.
\end{IEEEproof}

Now let us consider the factorization of a univariate constituent polynomial $Q_\nu(x)$, $\nu\in\mathcal{F}$, into a prefactor (re-encoding, Sierpinski, or combined) and the corresponding quotient polynomial ($U_\nu(x)$, $V_\nu(x)$, or $W_\nu(x)$). In order to prescind from the actual type of factorization let us denote the prefactor (whatever type it is) by $F_\nu(x)=\sum_{\mu=0}^{\deg[F_\nu(x)]} F_{\nu, \mu}x^\mu$ and the corresponding quotient polynomial by $G_\nu(x)=\sum_{\mu=0}^{\deg[G_\nu(x)]} G_{\nu, \mu}x^\mu$ with $\deg[G_\nu(x)]\leq\maxdeg{G}{\nu}$. This gives $Q_\nu(x)=G_\nu(x)F_\nu(x)$ for $\nu\in\mathcal{F}$ with coefficients
\begin{equation}\label{eqn:Qcoeff}
  Q_{\nu, \mu}=\sum_{i=0}^{\mu} G_{\nu, \mu-i} F_{\nu, i},\quad \mu=0, \ldots, \maxdeg{Q}{\nu},
\end{equation}
where we implicitly used that the $i$th coefficient of a polynomial with $i<0$ or $i>$ the degree of the polynomial is zero.

In order to simplify the following description, let us agree on trivial prefactors $F_\nu(x)=1$ for all $Q_\nu(x)$, $\nu\in\mathcal{F}^c$. In these cases, the quotient polynomials are $G_\nu(x)=Q_\nu(x)$ and $\maxdeg{G}{\nu}=\maxdeg{Q}{\nu}$.

Note that the constant term $F_{\nu, 0}$ of any prefactor is nonzero due to \eqref{eqn:Px} . This allows us to write
\begin{multline}\label{eqn:Gcoeff}
  G_{\nu, \mu} = \frac{Q_{\nu, \mu}-\sum_{i=1}^{\mu} G_{\nu, \mu-i} F_{\nu, i}}{F_{\nu, 0}}\\
    =\frac{Q_{\nu, \mu}}{F_{\nu, 0}} -\sum_{i=1}^\mu \frac{F_{\nu, i}G_{\nu, \mu-i}}{F_{\nu, 0}},\; \mu=0, \ldots, \maxdeg{Q}{\nu},
\end{multline}
which shows that $G_{\nu, \mu}$ is a linear combination of the $G_{\nu, i}$, $i=0, \ldots, \mu-1$, and $Q_{\nu, \mu}$.

What we will do in the following is to exploit \eqref{eqn:Gcoeff} for $\mu=0, \ldots, \maxdeg{G}{\nu}$ in order to obtain a solvable linear system whose solution comprises the coefficients of $G_\nu(x)$ (\emph{preparation}) and then to exploit \eqref{eqn:Qcoeff} for $\mu=\maxdeg{G}{\nu}+1, \ldots, \maxdeg{Q}{\nu}$ in order to dispose of the redundant columns of the coefficient matrix (\emph{reduction}). Both steps --- preparation and reduction --- are based on applying Lemma~\ref{lemma:lincomb_solution} with certain parameters. 

For simplicity, let us consider $\nu=0$. We can apply the lemma with $\beta_0=\nicefrac{1}{F_{0, 0}}$, and $\beta_1=\beta_2=\cdots=0$ in order to exchange solution variable $x_0=Q_{0, 0}$ for $y=G_{0, 0}$. After that, we can apply the lemma again with $\beta_0=\nicefrac{1}{F_{0, 0}}$, $\beta_{m-1}=-\nicefrac{F_{0, 1}}{F_{\nu, 0}}$, and $\beta_1=\cdots=\beta_{m-2}=0$ in order to exchange $x_0=Q_{0, 1}$ for $y=G_{0, 1}$. The process can be repeated until $Q_{0, \maxdeg{G}{0}}$ is replaced by $G_{0, \maxdeg{G}{0}}$, which closes the preparation step for $Q_0(x)$.

The reduction step is performed by applying Lemma~\ref{lemma:lincomb_solution} with
\begin{align*}
  \beta_{m-1-\maxdeg{G}{0}}&=F_{0, 1}\\
  \beta_{m-1-\maxdeg{G}{0}+1}&=F_{0, 2}\\
  \shortvdotswithin{=}
  \beta_{m-1}&=F_{0, \maxdeg{G}{0}+1}
\end{align*}
in order to exchange $x_0=Q_{0, \maxdeg{G}{0}+1}$ by $y=0$. Obviously, this allows to delete the last column from the coefficient matrix. The process is repeated with
\begin{align*}
  \beta_{m-1-\maxdeg{G}{0}}&=F_{0, 2}\\
  \beta_{m-1-\maxdeg{G}{0}+1}&=F_{0, 3}\\
  \shortvdotswithin{=}
  \beta_{m-1}&=F_{0, \maxdeg{G}{0}+2}
\end{align*}
in order to exchange $x_0=Q_{0, \maxdeg{G}{0}+2}$ by $y=0$ and (after deletion of the last column) repeated again until $x_0=Q_{0, \maxdeg{Q}{0}}$ is exchanged by $y=0$ (allowing to delete the last column), which happens for
\begin{align*}
  \beta_{m-1-\maxdeg{G}{0}}&=F_{0, \maxdeg{Q}{0}-\maxdeg{G}{0}}\\
  \beta_{m-1-\maxdeg{G}{0}+1}&=F_{0, \maxdeg{Q}{0}-(\maxdeg{G}{0}-1)}\\
  \shortvdotswithin{=}
  \beta_{m-1}&=F_{0, \maxdeg{Q}{0}}.
\end{align*}
Preparation and reduction are then executed for all remaining $\nu$, \ie $\nu=1, \ldots, \ell$. Algorithm~\ref{alg:reduction} cumulates the whole process.

As a result, the original coefficient matrix associated with \eqref{eqn:hasse}, whose $\sum_{\nu=0}^\ell (\maxdeg{Q}{\nu}+1)$ columns are associated with
\begin{equation*}
  Q_{0, 0}, \ldots, Q_{0, \maxdeg{Q}{0}}, Q_{1, 0}, \ldots, Q_{1, \maxdeg{Q}{1}}, \ldots, Q_{\ell, 0}, \ldots, Q_{\ell, \maxdeg{Q}{\ell}}
\end{equation*}
is converted into a reduced coefficient matrix, whose $\sum_{\nu=0}^{r-1} (\maxdeg{G}{\nu}+1)+\sum_{\nu=r}^\ell (\maxdeg{Q}{\nu}+1)$ columns are associated with
\begin{equation*}
  G_{0, 0}, \ldots, G_{0, \maxdeg{Q}{0}}, G_{1, 0}, \ldots, G_{1, \maxdeg{Q}{1}}, \ldots, G_{\ell, 0}, \ldots, G_{\ell, \maxdeg{Q}{\ell}}.
\end{equation*}

The quotient polynomials $G_\nu(x)$, $\nu\in\mathcal{F}$, as well as the $Q_\nu(x)$, $\nu\in\mathcal{F}^c$, can be directly read from any solution vector of the reduced linear system. The remaining $Q_\nu(x)$, $\nu\in\mathcal{F}$, can be reconstructed from the corresponding quotient polynomials and the prefactors $F_\nu(x)$. This allows to set up the bivariate polynomial $Q(x, z)=Q_0(x)+Q_1(x)+\cdots+Q_\ell(x)$ as a solution to the GSA interpolation step.

Note that Lemma~\ref{lemma:lincomb_solution} is formulated such that the leading column of the coefficient matrix $\vec{a}_0$ and the first solution variable $x_0$ is removed and a new column $\nicefrac{1}{\beta_0}\vec{a_0}$ and a new solution variable is appended. Thus, repeated application of the lemma processes the coefficient matrix from left to right, which makes Algorithm~\ref{alg:reduction} particularly easy to understand and at the same time saves a few indices in this part of the paper.

\begin{algorithm}[htbp]
\dontprintsemicolon
\vspace{0.5ex}

\SetKwFor{For}{Input:}{}{}
\For{}{
  $\GRSparloc$%
  \tgs*[f]{\algcomment{code parameters}}\;
  $m=\sum_{\nu=0}^\ell (\maxdeg{Q}{\nu}+1)$%
  \tgs*[f]{\algcomment{number of unknown coefficients}}\;
  $\mat{A}=\left(\vec{a}_0^T, \ldots, \vec{a}_{m-1}^T\right)$%
  \tgs*[f]{\algcomment{coefficient matrix, output of Algorithm~\ref{alg:GSA_matrix}}}\;
  $\mathcal{F}=\{\nu:0\leq\nu\leq\ell,Q_\nu(x)\;\text{has prefactor}\}$\;
  }
  
\SetKwFor{For}{For}{}{}

\For{$\nu$ from $0$ to $\ell$}
  {%
  \eIf{$\nu\in\mathcal{F}$}%
    {%
    calculate $F_\nu(x)$ and $\maxdeg{G}{\nu}$%
      \tgs*[f]{\algcomment{any type of prefactor}}\;
    }
    (\tgs*[f]{\algcomment{$\nu\in\mathcal{F}^c$}})
    {
    $F_\nu(x)\leftarrow 1$\;
    $\maxdeg{G}{\nu}\leftarrow\maxdeg{Q}{\nu}$\;
    }
  \For(\tgs*[f]{\algcomment{preparation}}){$\mu$ from $0$ to $\maxdeg{G}{\nu}$}
    {%
    \For{$j$ from $1$ to $\mu$}
      {%
      $\vec{a}_{m-1-j}\stackrel{+}{\leftarrow} F_j\vec{a}_0$\;
      }
    $\mat{A}\leftarrow \left(\vec{a}_1^T, \ldots, \vec{a}_{m-1}^T, F_0\vec{a}_0^T\right)$\;
    }
  \For(\tgs*[f]{\algcomment{reduction}}){$\mu$ from $\maxdeg{G}{\nu}+1$ to $\maxdeg{Q}{\nu}$}
    {%
    \For{$j$ from $1$ to $\maxdeg{G}{\nu}+1$}
      {%
      $\vec{a}_{m-1-j}\stackrel{+}{\leftarrow} F_{\mu-\maxdeg{G}{\nu}+j}\vec{a}_0$\;
      }
    $\mat{A}\leftarrow \left(\vec{a}_1^T, \ldots, \vec{a}_{m-1}^T\right)$\;
    $m\leftarrow m-1$\;
    }
  }

\vspace{0.5ex}
\KwOut{reduced coefficient matrix $\mat{A}$}
\caption{Reduction of GSA interpolation matrix.}
\label{alg:reduction}
\end{algorithm}

A rather technical analysis of the algorithm shows that columns $\vec{a}_i^T$ associated with $Q_{\nu, \mu}$, $\nu\in\mathcal{F}$ and $0\leq\mu\leq\maxdeg{G}{\nu}$, are replaced by $\sum_{\Delta=0}^{\maxdeg{Q}{\nu}-\maxdeg{G}{\nu}} F_{\nu, \Delta}\vec{a}_{i+\Delta}^T$ and associated with $G_{\nu, \mu}$, while columns associated with $Q_{\nu, \mu}$, $\nu\in\mathcal{F}$ and $\maxdeg{G}{\nu}<\mu\leq\maxdeg{Q}{\nu}$ are deleted. Columns associated with $Q_{\nu, \mu}$, $\nu\in\mathcal{F}^c$, are simply carried over from the original coefficient matrix. This allows the following reformulation of the GSA interpolation step:

\begin{problem}[Reduced GSA Interpolation Step]\label{prob:interpolation_reduced}
  Given a received vector $\vec{y}$ with prefactors $F_\nu(x)$, $\nu\in\mathcal{F}$, find a nonzero bivariate polynomial
  \begin{equation*}  
    \widetilde{Q}(x, z)=\sum_{\nu\in\mathcal{F}} G_\nu(x)y^\nu+\sum_{\nu\in\mathcal{F}^c} Q_\nu(x)y^\nu\in\Fq[x, z]
  \end{equation*}
  such that $\deg\left[G_\nu(x)\right]\leq\maxdeg{G}{\nu}$ and $\deg\left[Q_\nu(x)\right]\leq\maxdeg{Q}{\nu}$ and\vspace{-0.3cm}
  \begin{multline}\label{eqn:hasse_reduced}
    \forall i\in\mathcal{I}\;\forall s,t\in\N:s+t<r\;\text{and}\\
    \sum_{\substack{\nu\in\mathcal{F}\\\nu\geq t}}\binom{\nu}{t}z^{\nu-t}%
    \sum_{\mu=0}^{\maxdeg{G}{\nu}}%
    \underbrace{\left(\sum_{\Delta=0}^{\maxdeg{Q}{\nu}-\maxdeg{G}{\nu}}%
    \binom{\mu+\Delta}{s}x^{\mu+\Delta-s}F_{\nu, \Delta}\right)}_{\defeq\mathrm{LUT}[s, i, \nu, \mu]}%
    G_{\nu, \mu}\\
    +\sum_{\substack{\nu\in\mathcal{F}^c\\\nu\geq t}}\binom{\nu}{t}z^{\nu-t}%
    \sum_{\mu=s}^{\maxdeg{Q}{\nu}}
    \underbrace{\binom{\mu}{s}x^{\mu-s}}_{\defeq\mathrm{LUT}[s, i, \nu, \mu]}%
    Q_{\nu, \mu}
    \bigg|_{(x, z)=\left(\alpha^{-i}, y_i\right)}=0.
  \end{multline}
\end{problem}

Note that in case of re-encoding or combined prefactors with respect to $\mathcal{J}\subseteq\mathcal{I}$, $\vert\mathcal{J}\vert=k$, we can replace $\mathcal{I}$ by $\mathcal{J}$ in \eqref{eqn:hasse_reduced}. In that case, the input vector $\vec{y}$ must be the new received vector after the re-encoding projection, \ie $\vec{y}=\REENC{\mathcal{J}}{\vec{y}'}$ with the actual received vector $\vec{y}'$. The coefficient matrix associated with the reduced interpolation problem can be set up using Algorithm~\ref{alg:GSA_matrix_reduced}.

We emphasize that the sums in \eqref{eqn:hasse_reduced} with summation index $\mu$ are independent of the received vector $\vec{y}$ and thus their addends can be pre-calculated and stored in an $(r-1)\times n\times\ell\times \maxdeg{Q}{0}$ lookup table $\mathrm{LUT}[s, i, \nu, \mu]$.\footnote{The size of the lookup table is just a coarse upper bound, the memory requirements for the actual table are significantly smaller.} Hence, the overlapping double summations that occur for $\nu\in\mathcal{F}$ have no negative effect on the complexity of setting up the coefficient matrix.

As mentioned before, a solution $Q(x, z)$ of the GSA interpolation step (Problem~\ref{prob:interpolation}) can be recovered from a solution $\widetilde{Q}(x, z)$ of the reduced GSA interpolation step (Problem~\ref{prob:interpolation_reduced}) using the prefactors. $Q(x, z)$ can then be used as input to the GSA factorization step (Problem~\ref{prob:factorization}) in order to complete the decoding. A special case of the re-encoding projection that makes the reconstruction of $Q(x, z)$ particularly simple and hardware-friendly has been introduced in \cite{senger:2012,senger:2013a}. A \emph{reduced} GSA factorization step that could operate directly on $\widetilde{Q}(x, z)$ in order to construct the result list was proposed in \cite{koetter_ma_vardy_ahmed:2003,gross_kschischang_koetter_gulak:2005,ahmed_koetter_shanbhag:2011, koetter_ma_vardy:2011}. The focus of this paper is on interpolation, which is why we do not delve into the details of factorization.

\begin{example}\label{ex:matrix1}
For the setting of Example~\ref{ex:sier1}, Table~\ref{tab:sier_example} delivers a sum of univariate constituent degrees of $1421$, which means that the number of unknowns in the original GSA interpolation step is $1421+14=1435$. In contrast to this, we obtain $1227$ unknowns for the reduced GSA interpolation step presented in this section if only Sierpinski prefactors are used, $555$ unknowns if only re-encoding prefactors are used, and a mere $475$ unknowns if both types of prefactors are combined. The reduced coefficient matrices for the latter three cases can be set up using Algorithm~\ref{alg:GSA_matrix_reduced} and the following inputs:
\begin{enumerate}[(1)]
  \item Sierpinski: $\mathcal{J}=\emptyset$, $\mathcal{F}=\{5, 6, 7, 8\}$, $G_\nu(x)=V_\nu(x)$ and $F_\nu(x)=P_\mathcal{I}(x)^{10-g[\nu]}$ as in Theorem~\ref{thm:sier_factors}. 
  \item re-encoding: $\mathcal{J}=\{10, \ldots, 25\}$, $\mathcal{F}=\{0, \ldots, 9\}$, $G_\nu(x)=U_\nu(x)$ and $F_\nu(x)=P_\mathcal{J}(x)^{10-\nu}$ as in Theorem~\ref{thm:reenc_factors}.
  \item combined: $\mathcal{J}=\{10, \ldots, 25\}$, $\mathcal{F}=\{0, \ldots, 9\}$, either $G_\nu(x)=W_\nu(x)$, $F_\nu(x)=P_\mathcal{J}(x)^{10-\nu} P_{\mathcal{I}\setminus\mathcal{J}}(x)^{10-g[\nu]}$ or $G_\nu(x)=U_\nu(x)$, $F_\nu(x)=P_\mathcal{J}(x)^{10-\nu}$ as in Theorem~\ref{thm:combined_factors}.
\end{enumerate}
\end{example}

\begin{example}\label{ex:matrix2}
It turns out that Sierpinski prefactors work particularly well for the two conventional RS codes $\GRS_1(\F_{255}; 255, 191, 65)$ and $\GRS_2(\F_{255}; 255, 144, 112)$ considered by K\"otter and Vardy in \cite{koetter_vardy:2003}. The GSA for $\GRS_1$ can correct up to $\varepsilon_0=34$ errors with multiplicity $r=16$ and list size $\ell=18$. This requires solving a linear system in $34694$ unknowns. Sierpinski prefactors alone reduce the system to $31379$ unknowns, re-encoding alone to $8718$ unknowns. When both techniques are combined, the number of unknowns shrinks to $7886$.

A practically probably more relevant example is the GSA with multiplicity $r=4$ and list size $\ell=5$ for $\GRS_2$, it can correct up to $\varepsilon_0=59$ errors, which is four errors beyond half its minimum distance. The associated linear system has $2559$ unknowns, which can be diminished to $2049$ unknowns using Sierpinski prefactors alone, $1119$ unknowns using re-encoding alone, and a mere $897$ unknowns when both techniques are combined. This is only slightly more than the $676$ unknowns that are required to decode up to $\varepsilon_0=57$ errors with multiplicity $r=3$ and list size $\ell=4$ using re-encoding prefactors. Hence, the newly introduced combined prefactors (based on Sierpinski prefactors) allow to correct two additional errors at the comparatively low cost of having $221$ additional unknowns.
\end{example}

\section{Conclusion and Outlook}\label{sec:conclusion}

We introduced the concept of a-priori known prefactors in the GSA interpolation step and showed how the well-known re-encoding projection can be interpreted in this framework. Our main contribution is Section~\ref{sec:sierpinski}, where we introduced the new type of Sierpinski prefactors, which exist for a fairly wide range of code and GSA parameters. As opposing to re-encoding prefactors, Sierpinski prefactors do not require a modification of the received vector since they are based on a simple property of the ground field. In general, prefactors allow to reduce the linear system of equations that is associated with the GSA interpolation step, \ie they allow to reconstruct a solution of the original big system from the solution of a reduced smaller system. The reduction of the linear system for re-encoding prefactors, Sierpinski prefactors, and the (practically most relevant) combination of both has been described in Section~\ref{sec:factors}.

We stress that solving the reduced linear system using Gaussian elimination is straightforward with complexity cubic in $n$. However, it should not be too hard to adapt techniques like those from \cite{olshevsky_shokrollahi:2003,zeh_gentner_bossert:2009} for solving the reduced system with \emph{quadratic} complexity in $n$ since the structure of the original coefficient matrix is only marginally obstructed by the simple linear operator associated with the matrix reduction Algorithm~\ref{alg:reduction}. In similar manner, it should be possible to adapt K\"otter interpolation \cite{koetter:1996} such that arbitrary prefactors can be exploited. This would be a generalization of results from \cite{koetter_ma_vardy:2011}. We leave these problems for future work.

Another way to generalize our results would be to consider varying multiplicities for the received symbols, \eg in the context of algebraic soft-decision decoding with the K\"otter--Vardy algorithm \cite{koetter_vardy:2003}.

\appendices

\section{Setting up GSA Interpolation Matrices}%

Note that Lines~\ref{algline:GSA_matrix_reduced:betainit}, \ref{algline:GSA_matrix_reduced:deltaloop}, and \ref{algline:GSA_matrix_reduced:betaincr} in Algorithm~\ref{alg:GSA_matrix_reduced} can be deleted if $\beta$ in Line~\ref{algline:GSA_matrix_reduced:setval} is replaced by $\mathrm{LUT}[s, i, \nu, \mu]$ as in \eqref{eqn:hasse_reduced}. The algorithm admits a number of improvements that we ignored for the sake of slender pseudo code, \eg the binomial coefficient $\binom{\nu}{t}$ can be calculated directly at the beginning of the $\nu$-loop in Line~\ref{algline:GSA_matrix_reduced:nuloop} and the rest of the loop can be skipped in case it is zero modulo $p$.

Solving the linear system with a reduced coefficient matrix obtained by first calculating the full coefficient matrix with Algorithm~\ref{alg:GSA_matrix} and then reducing it with Algorithm~\ref{alg:reduction} results in the same solution (space) as directly solving the linear system with reduced coefficient matrix as obtained from Algorithm~\ref{alg:GSA_matrix_reduced}.

\begin{algorithm}[htbp]
\dontprintsemicolon
\vspace{0.5ex}

\SetKwFor{For}{Input:}{}{}
\For{}{
  $\GRSparloc$%
  \tgs*[f]{\algcomment{code parameters}}\;
  $r, \ell$%
  \tgs*[f]{\algcomment{multiplicity and list size}}\;
  }
  
\SetKwFor{For}{For}{}{}

$\mat{A}\leftarrow$ zero matrix over $\Fq$ with $n\binom{s+1}{2}$ rows\newline\hspace{1cm} and $\sum_{\nu=0}^\ell(\maxdeg{Q}{\nu}+1)$ columns\;
$o_\mathrm{r}\leftarrow  0$%
  \tgs*[f]{\algcomment{row offset}}\;
\For(\tgs*[f]{\algcomment{partial Hasse derivative in $z$}}){$s$ from $0$ to $r-1$}
  {%
  \For(\tgs*[f]{\algcomment{partial Hasse derivative in $x$}}){$t$ from $0$ to $r-s-1$}
    {%
    \For(\tgs*[f]{\algcomment{received symbol}}){$j$ from $0$ to $n-1$}
      {%
      $o_\mathrm{c}\leftarrow \sum_{\nu=0}^{t-1}(\maxdeg{Q}{\nu}+1)$%
        \tgs*[f]{\algcomment{column offset}}\;
      \For(\tgs*[f]{\algcomment{constituent polynomial $Q_\nu(x)$}}){$\nu$ from $t$ to $\ell$}
        {%
        \For(\tgs*[f]{\algcomment{coefficient of $Q_\nu(x)$}}){$\mu$ from $s$ to $\maxdeg{G}{\nu}$}
          {%
            $A_{o_\mathrm{r}+j, o_\mathrm{c}+\mu}\leftarrow%
              \binom{\nu}{t}\binom{\mu}{s}%
              \alpha_j^{\mu-s}%
              y_j^{\nu-t}$\;
          }
        $o_\mathrm{c}\stackrel{+}{\leftarrow} \maxdeg{Q}{\nu}+1$%
          \tgs*[f]{\algcomment{increment column offset}}\;
        }
      }
    $o_\mathrm{r}\stackrel{+}{\leftarrow} n$%
      \tgs*[f]{\algcomment{increment row offset}}\;
    }
  }

\vspace{0.5ex}
\KwOut{coefficient matrix $\mat{A}$}
\caption{GSA interpolation matrix.}
\label{alg:GSA_matrix}
\end{algorithm}

\begin{algorithm}[htbp]
\dontprintsemicolon

\vspace{0.5ex}

\SetKwFor{For}{Input:}{}{}
\For{}{
  $\GRSparloc$%
  \tgs*[f]{\algcomment{code parameters}}\;
  $r, \ell, \mathcal{J}$%
  \tgs*[f]{\algcomment{multiplicity and list size, re-encoding positions, $\vert\mathcal{J}\vert\in\{0, k\}$}}\;
  $\mathcal{F}=\{\nu:0\leq\nu\leq\ell,Q_\nu(x)\;\text{has prefactor}\}$\;
  }
  
\SetKwFor{For}{For}{}{}

\For(\tgs*[f]{\algcomment{calculate and store prefactors and degrees}}){$\nu$ from $0$ to $\ell$}
  {%
  \eIf{$\nu\in\mathcal{F}$}%
    {%
    calculate $F_\nu(x)$ and $\maxdeg{G}{\nu}$%
      \tgs*[f]{\algcomment{any type of prefactor}}\;
    }
    (\tgs*[f]{\algcomment{$\nu\in\mathcal{F}^c$}})
    {
    $F_\nu(x)\leftarrow 1$, $\maxdeg{G}{\nu}\leftarrow\maxdeg{Q}{\nu}$%
      \tgs*[f]{\algcomment{$G_\nu(x)=Q_\nu(x)$}}\;
    }   
  }
$\mat{A}\leftarrow$ zero matrix over $\Fq$ with $(n-\vert\mathcal{J}\vert)\binom{s+1}{2}$ rows\newline\hspace{1cm} and $\sum_{\nu=0}^\ell(\maxdeg{G}{\nu}+1)$ columns\;
$o_\mathrm{r}\leftarrow  0$%
  \tgs*[f]{\algcomment{row offset}}\;
\For(\tgs*[f]{\algcomment{partial Hasse derivative in $z$}}){$s$ from $0$ to $r-1$}
  {%
  \For(\tgs*[f]{\algcomment{partial Hasse derivative in $x$}}){$t$ from $0$ to $r-s-1$}
    {%
    $a\leftarrow 0$\;
    \For(\tgs*[f]{\algcomment{received symbol}}){$j$ from $0$ to $n-1$}
      {%
      \eIf(\tgs*[f]{\algcomment{skip re-encoding positions}}){$j\in\mathcal{J}$}
        {
        $a\stackrel{+}{\leftarrow} 1$\;
        }
        {
        $o_\mathrm{c}\leftarrow \sum_{\nu=0}^{t-1}(\maxdeg{G}{\nu}+1)$%
         \tgs*[f]{\algcomment{column offset}}\;
        \For(\tgs*[f]{\algcomment{quotient polynomial $G_\nu(x)$}}){$\nu$ from $t$ to $\ell$\nllabel{algline:GSA_matrix_reduced:nuloop}}
          {%
          $\beta\leftarrow 0$\;\nllabel{algline:GSA_matrix_reduced:betainit}
          \For(\tgs*[f]{\algcomment{coefficient of $G_\nu(x)$}}){$\mu$ from $0$ to $\maxdeg{G}{\nu}$}
            {%
            \For(\tgs*[f]{\algcomment{coefficient of $F_\nu(x)$}}){$\Delta$ from $0$ to $\maxdeg{Q}{\nu}-\maxdeg{G}{\nu}$\nllabel{algline:GSA_matrix_reduced:deltaloop}}
              {%
              $\beta\stackrel{+}{\leftarrow}%
                \binom{\mu+\Delta}{s}%
                F_{\nu, \Delta}\alpha_j^{\mu+\Delta-s}$\;\nllabel{algline:GSA_matrix_reduced:betaincr}
              }
            }
          $A_{o_\mathrm{r}+j-a, o_\mathrm{c}+\mu}\leftarrow\binom{\nu}{t}y_j^{\nu-t}\beta$\;\nllabel{algline:GSA_matrix_reduced:setval}
          $o_\mathrm{c}\stackrel{+}{\leftarrow} \maxdeg{G}{\nu}+1$%
            \tgs*[f]{\algcomment{increment column offset}}\;
          }
        }
      }
    $o_\mathrm{r}\stackrel{+}{\leftarrow} n-\vert\mathcal{J}\vert$%
      \tgs*[f]{\algcomment{increment row offset}}\;
    }
  }

\vspace{0.5ex}
\KwOut{reduced coefficient matrix $\mat{A}$}
\caption{Reduced GSA interpolation matrix.}
\label{alg:GSA_matrix_reduced}
\end{algorithm}

\section*{Acknowledgment}

The author is grateful to Frank R.\ Kschischang for stimulating discussions and valuable comments on the manuscript.

\ifCLASSOPTIONcaptionsoff
  \newpage
\fi


\end{document}